\documentclass[ajp,preprint,superscriptaddress,showpacs,preprintnumbers,eqsecnum,amsmath,amssymb]{revtex4}

\usepackage{graphicx}% Include figure files
\usepackage{dcolumn}% Align table columns on decimal point
\usepackage{bm}% bold math
\usepackage{color}
\usepackage{epsfig}
\usepackage{psfrag}
\usepackage{amsfonts}

\definecolor{darkgreen}{rgb}{0,0.6,0}

\newcommand{\comment}[1]{}
\input xy %to make diagrams
\xyoption{all}

\begin{document}

\title{
Classical and quantum-mechanical state reconstruction
}

\author{F. C. Khanna}
\affiliation 
{Theoretical Physics Institute, Physics Department, University of Alberta, Edmonton, Alberta, Canada T6G 2J1}

\author{P. A. Mello}
\affiliation
{Instituto de F\'isica, U.N.A.M., Apartado Postal 20-364, 01000 M\'exico, D. F.}

\author{M. Revzen}
\affiliation
{Department of Physics, Technion - Israel Institute of
Technology, Haifa 32000, Israel}

%%%%%%%%%%%%%%%%%%%%%%%%%%%%%%%%%
\begin{abstract}
We review the problem of state reconstruction in classical and in quantum physics, which is rarely considered at the textbook level.
We review a method for retrieving a classical state in phase space, similar to that used in medical imaging known as Computer Aided Tomography.
We explain how this method can be taken over to quantum mechanics, where it leads to a description of the quantum state in terms of the Wigner function which, although may take on negative values, plays the role of the probability density in phase space in classical physics.
We explain another approach to quantum state reconstruction based on the notion of Mutually Unbiased Bases, and indicate the relation between these two approaches.
Both are for a continuous, infinite-dimensional Hilbert space.
We then study the finite-dimensional case and show how the second method, based on Mutually Unbiased Bases, can be used for state reconstruction.
\end{abstract}
%%%%%%%%%%%%%%%%%%%%%%%%%%%%%%%%%%%%


\maketitle

%%%%%%%%%%%%%%%%%%%%%%%%%%%%%%%%%%%%%%%%%%%
\section{Introduction}
\label{intro}

The retrieval of the state of a physical system is an important problem in classical as well as in quantum physics, and yet it is a subject which is seldom discussed in textbooks.

The state of a system in {\em classical} physics is described by a density in  phase space, $\rho(q,p)$, which could be determined (i.e., reconstructed) by the directly measurable conditional probability density of the position $q$ for a given momentum $p$, $P(q|p)$, and the measurable probability density of $p$, $P(p)$, through the relation
%%%%%%%%%%%%%%%%%
\begin{equation}
\rho(q,p) = P(q|p) P(p)  .
\label{rho(q,p)}
\end{equation}
%%%%%%%%%%%%%%%%%%%
Thus we may envision the state as being specified by the set of measurable quantities $P(q|p)$ and $P(p)$.
An alternative approach for determining a classical state involves measuring
a linear combination of position and momentum (the constants $a$ and $b$ are introduced for the purpose of fixing dimensions)
%%%%%%%%%%%%%%%%%
\begin{equation}
X_{\theta} = a C q + b S p, \;\;\;\;\; C=\cos \theta, \;\; S=\sin \theta,
\label{Xtheta}
\end{equation}
%%%%%%%%%%%%%%%%%%%
sometimes termed, for electromagnetic-field state measurements, ``rotated quadratures" (Ref. \cite{schleich}, p. 136). 
The probability for the new variable $X_{\theta}$ for all values of $\theta$ can then be used to reconstruct the phase-space density $\rho(q,p)$ \cite{raymer}.
This procedure is similar to the familiar one employed in medical imaging for the reconstruction of a two-dimensional (2D) configuration density $\rho(x,y)$, known as the Computer Aided Tomography (CAT) scan \cite{guy,raymer,freeman}:
one simply replaces the two-dimensional configuration-space variables $(x,y)$ of the CAT method by the two-dimensional phase-space variables $(q,p)$.

In {\em quantum} physics, a system may be prepared in a pure state described by a vector in Hilbert space, or, more generally, in a mixed state described by a density operator $\hat{\rho}$
(Ref. \cite{messiah}, p. 204 and Ref. \cite{peres}, p. 72).
The problem of state retrieval involves the inverse inquiry, i.e., what are the measurable quantities whose values will suffice to determine the quantum state.
Historically, this question may be traced back to the Pauli query \cite{pauli} whether one can reconstruct the wave function, amplitude and phase,
for a one-particle system, from the probability of its position, i.e., 
$|\psi(x)|^2$, and that for its momentum, $|\tilde{\psi}(p)|^2$; 
here $\tilde{\psi}(p)$ is the wave function in the momentum representation, the tilde indicating the Fourier transform.
We now know that, in general, this is not possible: we need more information than these two distributions. 
The literature on this subject, which is still of current interest, has grown enormously ever since. 
Here we have made a selection out of these approaches, with the idea of providing a link with the classical reconstruction scheme.

The classical approach based on $P(q|p)$,
is, of course, untenable in quantum physics, where a fixed momentum precludes a well-defined position probability.
A similar observation is applicable to the direct approach of measuring the joint probability of $q$ and $p$.
However, it is remarkable that the alternative method based on measuring $X_{\theta}$
defined in phase space can be taken over to 
{\em quantum} mechanics (Ref. \cite{schleich}, p. 143, Ref. \cite{ulf}, p. 101).
But then the question arises: how can that be, if there is no such thing as a joint probability density $\rho(q,p)$ in quantum mechanics?
It turns out that the answer one obtains by following this procedure is a function defined in phase space which, although is not a bona-fide probability density 
(it is real, but not-necessarily non-negative, and has sometimes been named a ``quasi-probability"),
contains all the information needed to compute any quantum mechanical expectation value we please, just as if we were given the complex wave function, or the density operator.
This concept of quasi-probability was invented by Wigner \cite{wigner} in the early days of Quantum Mechanics, with the purpose of finding the quantum-mechanical corrections to thermodynamic functions, and is known as the
{\em Wigner function}.
Thus retrieving the Wigner function using this tomographic method is a true quantum-state reconstruction, and to explain how this is achieved,
and its relation with the classical tomographic approach, constitutes the main goal of the present paper.
The main results for this approach are to be found in 
Eq. (\ref{inverse radon q,p}) below for the classical case, and in 
Eq. (\ref{inverse radon q,p QM}) for the quantum-mechanical one.

There is another concept which has been very useful in the task of reconstructing a quantum state.
To give a trivial example, consider the eigenvectors of position and momentum: if the state vector of a system is an eigenstate of momentum, the system is 
{\em equally likely} to be found in any of the eigenstates of position.
Pairs of bases with a similar property have been extensively studied 
\cite{schwinger,wootters2} and are known as 
{\em Mutually Unbiased Bases} (MUB).
It turns out that MUB constitute a powerful tool for state reconstruction, since it is possible to express the density operator that defines the state of the system in terms of a complete orthonormal set of operators \cite{wootters3,filippov,gibbons_et_al_2004}.
We will explain the MUB approach to the problem of state reconstruction 
and show that the result [see Eq. (\ref{Wfctn of rho(MUB)}) below] is consistent with that found with the method explained above, based on tomography in phase space and the Wigner function.
Even more important, we shall find that the two approaches correspond, essentially, to employing two ways of handling the same complete set of operators, thus providing a unified description of both methods.

The paper is organized as follows.
In the next section we review the CAT scan method, as employed for the reconstruction of a classical 2D density.
In Section \ref{classical reconstruction} we present a scheme for classical state reconstruction in phase space similar to the CAT method employed in configuration space.
In Section \ref{quantum reconstruction} we explain how the classical scheme can be taken over to quantum mechanics, and explain the role played by Wigner function.
We then present in Sec. \ref{mub} the alternative method for quantum state reconstruction based on the notion of MUB.
So far, the discussion has been restricted to quantum systems described in a continuous, infinite-dimensional Hilbert space, because of our desire to make an analogy with classical physics. 
However, there have been many contributions to the problem of state reconstruction for quantum systems described in a finite-dimensional Hilbert space employing the notion of MUB.
Although these systems do not have a classical counterpart, still they allow us to draw an illuminating parallel with the various concepts that have been introduced for a continuous Hilbert space.
This fact motivates the brief discussion on the role of MUB for a finite-dimensional Hilbert space presented in Sec. \ref{finite mub} .
Finally, we give our conclusions in Sec. \ref{concl}.
To avoid cluttering of the main text, we include some details of the mathematical derivations in a number of appendices.

We wish to emphasize that the main goal for writing this paper
is to give a pedagogical presentation of a subject
which has been studied for many years and is still of current interest. 
With this motivation, we use a language that can be followed by a physics graduate student.
We do hope that the analysis is in a form that allows its incorporation in a graduate Quantum Mechanics course.

%%%%%%%%%%%%%%%%%%%%%%%%%%%%%%%%%%
\section{The classical reconstruction scheme}
\label{classical_reconstruction}

First we review briefly the method mentioned in the Introduction, 
the CAT scan, 
that is used for the reconstruction of a two-dimensional (2D) configuration density $\rho(x,y)$. The mathematical procedure can be translated directly to retrieve a classical 2D phase-space density $\rho(q,p)$ and, even more interesting for us, it can be taken over to quantum mechanics.

\begin{figure}[h]
\epsfxsize=0.63\textwidth
\epsfysize=0.4\textwidth  
\centerline{\epsffile{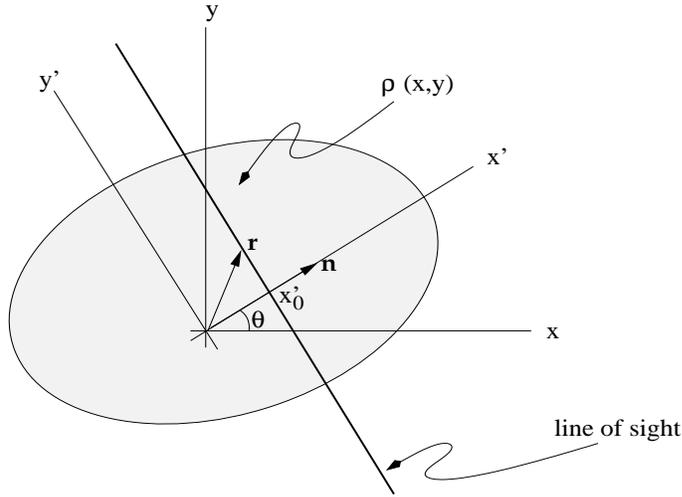}}
\caption{
\footnotesize{
In 2D Computer-Aided-Tomography (CAT), a beam of X-rays is passed through a sample, indicated by the shaded area, along the ``line of sight" which is offset by the amount $x'_0$ from the rotated $y'$ axis.
The unit vector ${\bf n}$, perpendicular to the line of sight, forms an angle $\theta$ with the $x$-axis.
Knowing the response of the sample for all offsets $x'$ and directions $\theta$ we can reconstruct the original sample density $\rho(x,y)$.
}}
\label{cat}
\end{figure}

In a 2D CAT-scan \cite{guy,raymer,freeman} a fine pencil beam of X-rays passes through a sample, shown as the shaded area in Fig. \ref{cat}, along the ``line of sight" defined by 
%%%%%%%%%%%%%%%%%%%%%
$
{\bf r \cdot n} = x'_0 ;
$
%%%%%%%%%%%%%%%%%%%%%%%%%
${\bf r}$ is the position vector of a point on the ``line of sight"
and ${\bf n}$ a unit vector perpendicular to the line of sight, forming an angle $\theta$ with the $x$-axis, so that ${\bf r}$ and ${\bf n}$ can be written as
%%%%%%%%%%%%%%%%%%%%%%%%%
%\begin{subequations}
%\begin{equation}
${\bf r} = x \; {\bf i} + y \; {\bf j}$,
${\bf n} = \cos  \theta \; {\bf i} + \sin \theta \; {\bf j}$,
%\end{equation}
%\end{subequations}
%%%%%%%%%%%%%%%%%%%%%%%%%
{\bf i} and {\bf j} being unit vectors along the $x$ and $y$ axes, respectively.
Then the equation for the line of sight becomes
%%%%%%%%%%%%%%%%%%%%%%%%%%%%%%%%%
\begin{equation}
x'_0 = x \cos \theta + y \sin \theta.
\label{line of sight b}
\end{equation}
%%%%%%%%%%%%%%%%%%%%%%%%%
The line of sight is offset by the amount $x'_0$ from the rotated $y'$ axis.
%%%%%%%%%%%%%%%%%%%%%%%%%%%%%%%%%%%%%%
%%%%%%%%%%%%%%%%%%%%%%%%%%%%%%%%%%%%%%
The beam is attenuated by scattering and absorption produced by 
the various parts of the sample encountered along the path.
Assuming that the attenuation at $(x,y)$ is proportional to the sample density $\rho(x,y)$, the total attenuation will be proportional to
%%%%%%%%%%%%%%%%%%%%%%%%
\begin{equation}
\rho_{\theta}(x')
= \int \int dxdy \; \delta(x'-Cx-Sy) \rho(x,y),
\label{radon}
\end{equation}
%%%%%%%%%%%%%%%%%%%%%%%%%%%%%
where we have used an arbitrary offset value designated by $x'$, and 
$C$ and $S$ have been defined in Eq. (\ref{Xtheta}). 
Now it is important to remark that knowing the response of the sample given by $\rho_{\theta}(x')$ for all $x'$ and directions $\theta$, we can reconstruct the density $\rho(x,y)$ of the sample.
The mathematics of this problem was actually developed by J. Radon at the beginning of the twentieth century \cite{radon} for the study of astronomical data. In fact, the function $\rho_{\theta}(x)$ of Eq. (\ref{radon}) is known in the literature as the {\em Radon transform} of the density $\rho(x,y)$. Thus the task is to invert the Radon transform to find the sample density.

It is shown in Appendix \ref{invert_radon} that the sample density $\rho(x,y)$ can be expressed in terms of the response of the sample, $\rho_{\theta}(x')$, for all $x'$ and directions $\theta$ defined above (see Ref. \cite{schleich}, pp. 144), as
%%%%%%%%%%%%%%%%%%%%%%%%
\begin{equation}
\rho(x,y)
= -\frac{1}{2 \pi^2} \int_0^{\pi} d \theta \;
{\cal P}\int_{-\infty}^{\infty} dx'
\frac{\partial \rho_{\theta}(x')/\partial x'}{x' - (x \cos \theta + y \sin \theta)} \; ,
\label{inverse radon}
\end{equation}
%%%%%%%%%%%%%%%%%%%%%%%%%%%%%
where ${\cal P}$ stands for the Cauchy principal value of the integral.
Indeed, Eq. (\ref{inverse radon}) is the inverse Radon transform of 
$\rho_{\theta}(x')$.

To gain some insight into the structure of the sample response $\rho_{\theta}(x')$, it is illustrative to consider the particular case in which the sample density $\rho(x,y)$
is isotropic, i.e., dependent only on the distance $r=\sqrt{x^2 + y^2}$ from the origin and independent of the angle. 
If we write $x$ and $y$ in polar coordinates as
$x=r\cos\phi$, $y=r\sin\phi$,
Eq. (\ref{radon}) for the response $\rho_{\theta}(x')$ takes the form
%%%%%%%%%%%%%%%%%%%%%%%%
\begin{eqnarray}
\rho_{\theta}(x')
&=& \int_0^{2\pi}d\phi \int_0^{\infty} dr \; r \; \delta(x'-r\cos(\phi - \theta)) \rho(r),
\nonumber \\
&=& \int_0^{2\pi}d\phi \int_0^{\infty} dr \; r \; \delta(x'-r\cos\phi) \rho(r),
\label{radon_isotropic}
\end{eqnarray}
%%%%%%%%%%%%%%%%%%%%%%%%%%%%%
showing that $\rho_{\theta}(x')$ is independent of $\theta$ for the isotropic case. 
By direct substitution, one may also observe that in this case $\rho_{\theta}(x')$ is symmetric, i.e., 
$\rho_{\theta}(-x') = \rho_{\theta}(x')$.

%%%%%%%%%%%%%%%%%%%%%%%%%%%%%%%%%%%%%%
\begin{figure}[h]
\epsfxsize=0.4\textwidth
\epsfysize=0.4\textwidth  
\centerline{\epsffile{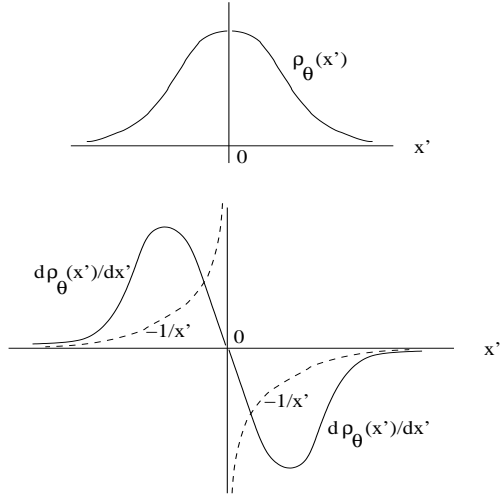}}
\caption{Schematic representation of the functions $\rho_{\theta}(x ')$, 
$d\rho_{\theta}(x')/d x'$ (solid lines) and $-1/x'$ (dashed line) as functions of $x'$, needed in the text to show that the sample density $\rho(0,0)$ at the origin is positive.
\footnotesize{
}}
\label{rho'(x')}
\end{figure}
%%%%%%%%%%%%%%%%%%%%%%%%%%%%%%%%%%%%%%
It may also be pointed out that the sample density $\rho(x,y)$ must be a
{\em non-negative} quantity, although this fact is not explicitly manifest in Eq. (\ref{inverse radon}).
It is thus useful to verify this property in some particular example.
For this purpose, we choose the isotropic case studied in the last paragraph.
For example, at the origin of coordinates, $x=y=0$, Eq. (\ref{inverse radon}) gives
%%%%%%%%%%%%%%%%%%%%%%%%
\begin{equation}
\rho(0,0)
= -\frac{1}{2 \pi} 
{\cal P}\int_{-\infty}^{\infty} dx'
\frac{\partial \rho_{\theta}(x')/\partial x'}{x'} \; ,
\label{inverse radon 1}
\end{equation}
%%%%%%%%%%%%%%%%%%%%%%%%%%%%%
Since $\rho_{\theta}(x ')$ is a symmetric function of $x'$, 
$\partial \rho_{\theta}(x')/\partial x'$ is antisymmetric. 
Since the quantity $-x'$ appearing in 
Eq. (\ref{inverse radon 1}) has precisely this same property, as illustrated in 
Fig. \ref{rho'(x')}, the resulting density $\rho(0,0)$ at the origin of coordinates is positive.

%%%%%%%%%%%%%%%%%%%%%%%%%%%%%%%%%%%%%
\section{Classical-Quantum physics state-reconstruction analogy}
\label{classical-quantum}

\subsection{Classical state reconstruction}
\label{classical reconstruction}

A state in {\em classical} statistical physics is determined by a probability density in phase space.
In this paper we shall always consider, for simplicity, 
one-particle systems with one degree of freedom.
We write the probability density in phase space as $\rho(q,p)$ which, for convenience in our comparison with Quantum Mechanics, will be normalized as
%%%%%%%%%%%%%%%%%%%%%%%%%%%
\begin{equation}
\int \rho(q,p) \frac{dq dp}{2\pi} 
= 1.
\label{int rho CL}
\end{equation}
%%%%%%%%%%%%%%%%%%%%%%%%%%%

After the discussion given in the previous section on CAT in 2D configuration space $(x,y)$, it is clear that a similar method can be applied in 2D phase space $(q,p)$:
if we consider the linear combination of position and momentum given in 
Eq. (\ref{Xtheta}),
the probabilities for the new variable $X_{\theta}$ for all values of $\theta$ can then be used to reconstruct $\rho(q,p)$ \cite{raymer}.

Before proceeding, we indicate our choice for the constants $a$ and $b$ which were introduced to fix dimensions.
We choose
%%%%%%%%%%%%%%%%%%%%%%%%%%%%%%%
$a=1/q_0$, $b=1/p_0$,
%%%%%%%%%%%%%%%%%%%%%%%%%%%
where $q_0$ and $p_0$ represent any convenient scales for position and momentum.
Subsequently renaming the {\em dimensionless} quantities $q/q_0$ and $p/p_0$ again as $q$ and $p$, respectively, the transformation of 
Eq. (\ref{Xtheta}) reads 
%%%%%%%%%%%%%%%%%%%%%%%%%%%%%%%
\begin{equation}
X_{\theta} = Cq + Sp.
\label{Xtheta 1 CM}
\end{equation}
%%%%%%%%%%%%%%%%%%%%%%%%%%%

We go back to the probability density of the variable $X_{\theta}$.
If we designate it as 
$\rho_{\theta}(x')$, where $x'$ represents an arbitrary value of $X_{\theta}$,
we have, just as in Eq. (\ref{radon})
%%%%%%%%%%%%%%%%%%%%%%%%%%%%%%
\begin{equation}
\rho_{\theta}(x')
= \int \int \; \delta(x'-Cq-Sp) \; \rho(q,p) \; \frac{dq dp}{2\pi} \; .
\label{rho(x')_vs_rho{q,p}}
\end{equation}
%%%%%%%%%%%%%%%%%%%%%%%%%%%
The goal is to find $\rho(q,p)$ in terms of $\rho_{\theta}(x')$ by inverting Eq. (\ref{rho(x')_vs_rho{q,p}}).
Proceeding as in the previous section and Appendix \ref{invert_radon}, we find the equivalent of Eq. (\ref{inverse radon}) as
%%%%%%%%%%%%%%%%%%%%%%%%
\begin{equation}
\rho(q,p)
= %|ab| \; 2 \pi \hbar \; 
-\frac{1}{\pi}\int_0^{\pi} d \theta \;
{\cal P}\int_{-\infty}^{\infty} dx'
\frac{\partial \rho_{\theta}(x')/\partial x'}
{x' - (q \cos \theta + p \sin \theta)} \; .
\label{inverse radon q,p}
\end{equation}
%%%%%%%%%%%%%%%%%%%%%%%%%%%%%

%%%%%%%%%%%%%%%%%%%%%%%%%%%%%%%%%%%%%%%%%%%%%%%%%%%%%%%%%%%%
\subsection{Quantum state reconstruction}
\label{quantum reconstruction}

As mentioned in the Introduction, the above method based on measuring $X_{\theta}$ defined in phase space can be taken over to 
{\em quantum} mechanics (Ref. \cite{schleich}, p. 143).
This leads to a {\em quasi-probability density} defined in phase space known as the Wigner function.

In what follows we shall take units in which $\hbar = 1$.
Consider an arbitrary Hermitean operator $\hat{A}$. We define its 
{\em Wigner transform} as \cite{ulf,wigner,schleich}
%%%%%%%%%%%%%%%%%%%%%%%%%%%%%%%%%
\begin{equation}
W_{\hat{A}}(q,p)
=\int e^{-ipy}
\left\langle q+\frac{y}{2} \left\vert\hat{A}\right\vert|q-\frac{y}{2}\right\rangle dy \; .
\label{WT}
\end{equation}
%%%%%%%%%%%%%%%%%%%%%%%%%%%%%%%%%% 
For the case where the operator $\hat{A}$ is the density operator $\hat{\rho}$ defining the state of the system, we speak of the 
{\em Wigner function} of the state, which has the normalization property
%%%%%%%%%%%%%%%%%%%%%%%%%%%%%%%%%
\begin{equation}
\int \int W_{\hat{\rho}}(q,p) \frac{dq dp}{2\pi}
=1 \; ,
\label{normal-WFrho}
\end{equation}
%%%%%%%%%%%%%%%%%%%%%%%%%%%%%%%%%%
similar to the normalization of Eq. (\ref{int rho CL}) adopted for the classical distribution.

It is well known \cite{schleich} that Wigner function for a state 
may be negative in some parts of phase space.
Thus it does not qualify as a true probability density and is referred to as a quasi-probability density. 
An illustration of the fact that it plays in quantum mechanics a role analogous to that played by the classical probability density $\rho(q,p)$ is the similarity of Eqs. (\ref{rhoQM(x')_WT 2}) and 
(\ref{inverse radon q,p QM}) given below with 
Eqs. (\ref{rho(x')_vs_rho{q,p}}) and (\ref{inverse radon q,p}), respectively.

An important property of Wigner function, obtained from the definition (\ref{WT}), is \cite{schleich,ulf}
%%%%%%%%%%%%%%%%%%%%%%%%%%%%%%%%%
\begin{equation}
Tr(\hat{A} \hat{B})
=\int \int W_{\hat{A}}(q,p) W_{\hat{B}}(q,p) \frac{dq dp}{2\pi} \; ,
\label{Tr(AB)}
\end{equation}
%%%%%%%%%%%%%%%%%%%%%%%%%%%%%%%%%%
for any two operators $\hat{A}$ and $\hat{B}$. 
This implies that the trace of the product of two operators in Hilbert space can be evaluated as an integral in phase space of the corresponding Wigner transforms.
The normalization of Eq. (\ref{normal-WFrho}) is consistent with the property (\ref{Tr(AB)}), taking 
$\hat{A}=\hat{\rho}$ and $\hat{B}=1$.
The statistical expectation value of an observable $\hat{A}$, obtained
by using Eq. (\ref{Tr(AB)}), can be expressed as
%%%%%%%%%%%%%%%%%%%%%%%%%%%%%%%%%
\begin{equation}
\langle \hat{A} \rangle
= Tr(\hat{\rho} \hat{A})
=\int \int W_{\hat{\rho}}(q,p) W_{\hat{A}}(q,p) \frac{dq dp}{2\pi} \; ,
\label{<A>}
\end{equation}
%%%%%%%%%%%%%%%%%%%%%%%%%%%%%%%%%%
i.e., as an integral in phase space of Wigner function for the state times Wigner transform of the observable.
With these results, Wigner function of the state and the Wigner transform of observables can be employed to ``do QM in phase space".

It is also a simple exercise to show that the above definition of the Wigner function of the state $\hat{\rho}$ is equivalent to the inverse Fourier transform of the characteristic function of the density operator
(Ref. \cite{ulf}, Eqs. (3.12), (3.16)), i.e., 
%%%%%%%%%%%%%%%%%
\begin{subequations}
\begin{eqnarray}
W_{\hat{\rho}}(q,p)
&=&\frac{1}{2\pi} \int \int \tilde{W}(u,v) e^{i(uq + vp)} du dv \; ,
\label{WT=invFT-char_fctn}
\\
\tilde{W}(u,v)&=&Tr[\hat{\rho} \; e^{-i(u\hat{q} + v \hat{p})}]  \; .
\label{char_fctn_rho}
\end{eqnarray}
\label{WT=invFT and char_fctn}
\end{subequations}
%%%%%%%%%%%%%%%%%%%

Now consider the observable
%%%%%%%%%%%%%%%%%
\begin{equation}
\hat{X}_{\theta} = C \hat{q} + S \hat{p}, 
\label{Xtheta_QM}
\end{equation}
%%%%%%%%%%%%%%%%%%%
which is the QM counterpart of the classical quantity of 
Eq. (\ref{Xtheta 1 CM}).
This observable satisfies the eigenvalue equation
%%%%%%%%%%%%%%%%%
\begin{equation}
\hat{X}_{\theta}|x'; \theta\rangle = x' |x'; \theta\rangle \; ,
\label{e-value eqn for Xtheta_QM}
\end{equation}
%%%%%%%%%%%%%%%%%%%
where $x'$ denotes an eigenvalue and $|x'; \theta\rangle$ the corresponding eigenvector.

Our program is as follows.
If the system is prepared in the state defined by the density operator $\hat{\rho}$, we first consider the probability density
$\rho^{QM}_{\theta}(x')$ that a measurement of the observable $\hat{X}_{\theta}$ will give the value $x'$: this probability density will be initially expressed in terms of $\hat{\rho}$, Eq. (\ref{rhoQM(x')}) below, and then in terms of the Wigner function $W_{\hat{\rho}}(q,p)$ in phase space, Eq. (\ref{rhoQM(x')_WT 2}) below. 
The final goal is to ``invert" this relation and show that we can retrieve the Wigner function in terms of $\rho^{QM}_{\theta}(x')$.

The probability density $\rho^{QM}_{\theta}(x')$ is given by the standard QM expression
%%%%%%%%%%%%%%%%%
\begin{equation}
\rho^{QM}_{\theta}(x')
= Tr(\hat{\rho} \; \mathbb{P}_{x'}^{\theta}) \; , 
\;\;\;{\rm where}\;\;\;
\mathbb{P}_{x'}^{\theta}
= |x'; \theta\rangle \;\langle x'; \theta |.
\label{rhoQM(x')}
\end{equation}
%%%%%%%%%%%%%%%%%%
Making use of Eq. (\ref{<A>}), we write 
%%%%%%%%%%%%%%%%%%%%%%%%%%%%%%%%%
\begin{equation}
\rho^{QM}_{\theta}(x')
=\int \int W_{\hat{\rho}}(q,p) W_{\mathbb{P}_{x'}^{\theta}}(q,p) 
\frac{dq dp}{2\pi} \; .
\label{rhoQM(x')_WT 1}
\end{equation}
%%%%%%%%%%%%%%%%%%%%%%%%%%%%%%%%%%
In this expression, $W_{\mathbb{P}_{x'}^{\theta}}(q,p)$ is the Wigner transform of the projector $\mathbb{P}_{x'}^{\theta}$, which is calculated in Appendix \ref{WT of Projector(theta)} with the result
%%%%%%%%%%%%%%%%%%%%%%%%%%%%%%%%%
\begin{equation}
W_{\mathbb{P}_{x'}^{\theta}}(q,p)
=\delta(x' - (C q + S p)) \; .
\label{WT_P_x'_theta}
\end{equation}
%%%%%%%%%%%%%%%%%%%%%%%%%%%%%%%%%%
Then Eq. (\ref{rhoQM(x')_WT 1}) takes the form
%%%%%%%%%%%%%%%%%%%%%%%%%%%%%%%%%
\begin{equation}
\rho^{QM}_{\theta}(x')
=\int \int W_{\hat{\rho}}(q,p) \; \delta(x' - (C q + S p)) \;
\frac{dq dp}{2\pi} \; .
\label{rhoQM(x')_WT 2}
\end{equation}
%%%%%%%%%%%%%%%%%%%%%%%%%%%%%%%%%%

This last equation is the QM counterpart of Eq. (\ref{rho(x')_vs_rho{q,p}}) for the classical probability density $\rho_{\theta}(x')$.
It shows explicitly that what plays the role of the classical probability density $\rho(q,p)$ in phase space is now the quasi-probability density given by the Wigner function $W_{\hat{\rho}}(q,p)$.
Thus in order to invert Eq. (\ref{rhoQM(x')_WT 2}) we just copy the result in Eq. (\ref{inverse radon q,p}) and write
$W_{\hat{\rho}}(q,p)$ in terms of $\rho^{QM}_{\theta}(x')$ as
%%%%%%%%%%%%%%%%%%%%%%%%
\begin{equation}
W_{\hat{\rho}}(q,p)
= - \frac{1}{\pi}\int_0^{\pi} d \theta \;
{\cal P}\int_{-\infty}^{\infty} dx'
\frac{\partial \rho_{\theta}^{QM}(x')/\partial x'}
{x' - (q \cos \theta + p \sin \theta)} \; .
\label{inverse radon q,p QM}
\end{equation}
%%%%%%%%%%%%%%%%%%%%%%%%%%%%%

This equation allows reconstructing the QM state, in the sense that from the observable probability density $\rho_{\theta}^{QM}(x')$ the Wigner function of the density operator can be extracted; its knowldedge, in turn,
is equivalent to that of the state itself.

This completes our analysis that shows a close analogy between the
classical and quantum state reconstruction: 
both require the use of the inverse Radon transform. We now turn to an alternative quantum-state reconstruction scheme which does not require the use of the Radon transform.

%%%%%%%%%%%%%%%%%%%%%%%%%%%%%%%%%%%
\section{Mutually Unbiased Bases and State Reconstruction}
\label{mub}

Mutually unbiased bases (MUB) in concept were introduced by Schwinger \cite{schwinger} in his studies of vectorial bases for Hilbert spaces that exhibit ``maximal degree of incompatibility''. 
The  eigenvectors of $\hat{x}$  and $\hat{p}$, $|x\rangle$ and
$|p\rangle$, respectively, are example of such bases.
The information-theoretical oriented appellation ``mutual unbiased
bases'' was introduced by Wootters \cite{wootters2}. 

Consider two complete and orthonormal vectorial bases, 
${\cal B}_1$, ${\cal B}_2$,
whose vectors will be designated by
$|u;{\cal B}_1 \rangle$ and $|v;{\cal B}_2 \rangle$, respectively.
The two bases are said to be MUB if and only if,
for ${\cal B}_1\ne {\cal B}_2$,
%%%%%%%%%%%%%%%%%%%%%%%%%%%%%
\begin{equation}
\left\vert \langle u;{\cal B}_1 | v;{\cal B}_2 \rangle \right\vert^2 
= K \; ,
\;\;\;\;\;
\forall u, v,
\label{MUB}
\end{equation}
%%%%%%%%%%%%%%%%%%%%%%%%%%%%%%%%
where $K$ is a constant {\em independent} of $u$ and $v$ 
(see Ref. \cite{filippov}).
This property means that the absolute value of the scalar product of vectors from {\em different} bases is independent of the vectorial label within either basis.
This implies that if a system is measured to be in one of the states,
say $|u; {\cal B}_1 \rangle$, of ${\cal B}_1$, it is equally likely to be found in any of the states $|v; {\cal B}_2\rangle$ of any other basis 
${\cal B}_2$, when ${\cal B}_1$ and ${\cal B}_2$ are MUB.
The value of $K$ may depend on the bases ${\cal B}_1$, ${\cal B}_2$, which indeed is the case for a continuous Hilbert space.
For a Hilbert space with a finite dimensionality $d$, $K=1/d$.

The concept of MUB is found to be of interest in several fields. 
For instance, the ideas are useful in a variety of cryptographic protocols \cite{ekert} and signal analysis \cite{vourdas}. 
%The present work utilizes the relation of MUB to quadrature observations %used \cite{schleich,wootters,ulf} in the reconstruction of Wigner %functions of quantum states.

In what follows we outline a scheme for state reconstruction based on MUB \cite{bengtsson} which is an alternative to the one presented in the previous section. 

%%%%%%%%%%%%%%%%%%%%%%%%%%%%%%%%%%
\subsection{Some properties of the operator $\hat{X}_{\theta}$ and its eigenstates}
\label{X-theta-and-its-e-states}

We begin with a review of the properties of the operator $\hat{X}_{\theta}$ and its eigenstates $|x',\theta \rangle$, Eqs. (\ref{Xtheta_QM}),
(\ref{e-value eqn for Xtheta_QM}), and show that the bases
$\{|x_1;\theta_1\rangle \}$, $\{|x_2;\theta_2\rangle\}$
($\theta_1 \neq \theta_2$, fixed) are MUB.

We repeat the definition (\ref{Xtheta_QM}) of the operator $\hat{X}_{\theta}$ and introduce the new operator $\hat{P}_{\theta}$ as
%%%%%%%%%%%%%%%%%%%%%%%%%%%%%%%%%%%%%%%%
\begin{equation}
\hat{X}_{\theta} = C \hat{x} + S \hat{p},
\;\;\;\;
\hat{P}_{\theta} = - S \hat{x} + C \hat{p} 
\; ;
\label{Xtheta,Ptheta}
\end{equation}
%%%%%%%%%%%%%%%%%%%%%%%%%%%%%%%%%%%%%%%%
$\hat{X}_{\theta}$ and $\hat{P}_{\theta}$ 
are {\em canonically conjugate}, i.e.,
%%%%%%%%%%%%%%%%%%%%%%%%%%%%%%%%
$
[ \hat{X}_{\theta}, \hat{P}_{\theta}] = i
$,
%%%%%%%%%%%%%%%%%%%%%%%%%%%%%%%%
just as the original operators $\hat{x}$, $\hat{p}$.

As a first step we solve the eigenvalue equation 
(\ref{e-value eqn for Xtheta_QM}) in the coordinate representation. In this representation we define the wave function
%%%%%%%%%%%%%%%%%%%%%%%%%%%%%%%%%
\begin{equation}
\psi_{x', \theta}(x)
= \langle x | x'; \theta \rangle,
\label{psi_{x,theta}(x')}
\end{equation}
%%%%%%%%%%%%%%%%%%%%%%%%%%%%%%%%
which satisfies the equation
%%%%%%%%%%%%%%%%%%%%%%%%%%%%%%%%%
\begin{equation}
\left(x \cos \theta - i \sin {\theta} \; \frac{\partial}{\partial x}      \right) \psi_{x', \theta}(x)
= x' \; \psi_{x', \theta}(x) .
\label{e-value eqn for psi_{x,theta}(x')}
\end{equation}
%%%%%%%%%%%%%%%%%%%%%%%%%%%%%%%%
The solution of this equation is
%%%%%%%%%%%%%%%%%%%%%%%%%%%%%%%%%
\begin{equation}
\psi_{x', \theta}(x)
= F(x', \theta) \; e^{-\frac{i}{2 \sin \theta}(x^2\cos \theta - 2x x')},
\label{psi_{x,theta}(x') 1}
\end{equation}
%%%%%%%%%%%%%%%%%%%%%%%%%%%%%%%%
where $F(x', \theta)$ is an arbitrary function of $x'$ and $\theta$.
It is shown in Appendix \ref{finding psi_x,theta(x')} that $F(x', \theta)$ can be completely determined, up to an arbitrary overall phase, by imposing on the states 
$|x'; \theta \rangle$ the requirements \cite{moshinsky}
%%%%%%%%%%%%%%%%%%%%%%%%%%%%%%%%%%%%%%%%
\begin{subequations}
\begin{eqnarray}
&&\langle x_1, \theta | x_2, \theta \rangle
=\delta (x_1 - x_2) ,
\label{orthonormal}
\\
\langle x_1, \theta |\hat{X}_{\theta}| x_2, \theta \rangle
&=& x_2 \; \delta (x_1 - x_2);
\hspace{5mm}
\langle x_1, \theta |\hat{P}_{\theta}| x_2, \theta \rangle
= -i \; \delta' (x_1 - x_2),
\label{m.els. X and P}
\\
\psi_{x', \theta}(x) = \langle x|x', \theta \rangle
&\to& \delta (x - x'), \;\;\; {\rm as} \;\;\; \theta \to 0;
\hspace{5mm} 
\psi_{x', \theta = \pi/2}(x) = \frac{e^{ix'x}}{\sqrt{2 \pi}} \; .
\label{lim theta to 0 and plane_wave}
\end{eqnarray}
\label{norm,m.els.X,P,theta to 0,plane_wave}
\end{subequations}
%%%%%%%%%%%%%%%%%%%%%%%%%%%%%%%

Here, Eq. (\ref{orthonormal}) expresses the ortho-normalization (in the sense of the Dirac delta function) of the states $|x'; \theta \rangle$. 
Equation (\ref{m.els. X and P}) requires that the matrix elements of the new canonically conjugate operators $\hat{X}_{\theta}$ and $\hat{P}_{\theta}$ with respect to the new states $|x'; \theta \rangle$ be equal to the matrix elements of the old canonically conjugate operators $\hat{x}$ and $\hat{p}$ with respect to the old states $|x \rangle$, as demanded by a {\em canonical transformation}.
The first Eq. (\ref{lim theta to 0 and plane_wave}) requires that in the limit  $\theta \to 0$ the overlap between the new state 
$|x', \theta \rangle$ and the old one $| x\rangle$ be a Dirac delta function.
The second Eq. (\ref{lim theta to 0 and plane_wave})
requires that for $\theta = \pi/2$, i.e., when 
$\hat{X}_{\theta = \pi/2}$ is the momentum $\hat{p}$, 
the wave function $\psi_{x', \theta = \pi/2}(x)$ be a plane wave with no extra phases.

The final result for the wave function $\psi_{x', \theta}(x)$, up to an overall constant phase, is
%%%%%%%%%%%%%%%%%%%%%%%%%%%%%%%%%
\begin{equation}
\psi_{x', \theta}(x)
= \frac{e^{i\left[\frac{\pi}{4}{\rm sgn}(\sin\theta)-\frac{\theta}{2}\right]}}{\sqrt{2 \pi |\sin \theta|}} \;
e^{-\frac{i}{2 \sin \theta}[(x'^2+x^2)\cos \theta - 2x x']} \; .
\label{psi_{x,theta}(x') 2}
\end{equation}
%%%%%%%%%%%%%%%%%%%%%%%%%%%%%%%%
Notice the symmetry of this expression under the interchange 
$x \leftrightarrow x'$.
Since $|-x', \theta + \pi \rangle = |x', \theta \rangle$, it suffices to consider state vectors 
in the range $-\infty < x' < \infty$ and $-\pi /2 \leq \theta \leq \pi /2$,
other values of $\theta$ repeating the eigenvectors in this range
(see Ref. \cite{schleich}, p. 144).

We relate the new state $|x', \theta \rangle$ to the old one 
$| x'\rangle$ through a {\em unitary transformation} as
%%%%%%%%%%%%%%%%%%%%%%%%%%%%%%%%%
\begin{equation}
|x', \theta \rangle = \hat{U}^{\dagger}(\theta) | x'\rangle.
\label{U 1}
\end{equation}
%%%%%%%%%%%%%%%%%%%%%%%%%%%%%%%%
For the reader's convenience, we mention that the operator $\hat{U}$ used in this article coincides with the one designated by $\hat{V}$ in 
Ref. \cite{micha-et-al-2005}, and that called $\hat{U}^{\dagger}$ in Ref. \cite{moshinsky}.
Using Eqs. (\ref{U 1}) and (\ref{psi_{x,theta}(x') 2}) we find, for the matrix elements of the unitary operator $\hat{U}^{\dagger}(\theta)$ in the old basis,
%%%%%%%%%%%%%%%%%%%%%%%%%%%%%%%%%%%%%%%%
\begin{equation}
\langle x|\hat{U}^{\dagger}(\theta)|x'\rangle
=\frac{e^{i\left[\frac{\pi}{4}{\rm sgn}(\sin \theta)
-\frac{\theta}{2}\right]}}{\sqrt{2 \pi |\sin \theta|}} \;
e^{-\frac{i}{2 \sin \theta}[(x'^2+x^2)\cos \theta - 2x x']} .
\label{<x'U+x>}
\end{equation}
%%%%%%%%%%%%%%%%%%%%%%%%%%%%%%%%%%%%%%%%
Using the unitary operator $\hat{U}(\theta)$ we write the eigenvalue equation (\ref{e-value eqn for Xtheta_QM}) as
%%%%%%%%%%%%%%%%%%%%%%%%%%%%%%%%%%%%%%%%
$
%\begin{equation}
\hat{U}(\theta) \hat{X}_{\theta}\hat{U}^{\dagger}(\theta) 
| x' \rangle
= x' | x' \rangle
%\label{e-value eqn for Xtheta_QM 1} \; 
%\end{equation}
$,
%%%%%%%%%%%%%%%%%%%%%%%%%%%%%%%%
implying
%%%%%%%%%%%%%%%%%%%%%%%%%%%%%%%%%%%%%%%%
$
\hat{x} 
= \hat{U}(\theta) \hat{X}_{\theta}\hat{U}^{\dagger}(\theta)
$.
%%%%%%%%%%%%%%%%%%%%%%%%%%%%%%%%%%%%%%%%
Thus the operator $\hat{x}$ and, similarly, its canonically conjugate $\hat{p}$ transform as
%%%%%%%%%%%%%%%%%%%%%%%%%%%%%%%%%%%%%%%%
\begin{equation}
\hat{X}_{\theta} 
=\hat{U}^{\dagger}(\theta) \; \hat{x} \; \hat{U}\;(\theta), \;\;\;\;
{\rm and} \;\;\;\;
\hat{P}_{\theta} 
= \hat{U}^{\dagger}(\theta) \; \hat{p} \; \hat{U}\;(\theta).
\label{X=UxU,P=UpU}
\end{equation}
%%%%%%%%%%%%%%%%%%%%%%%%%%%%%%%%
The above unitary transformation is given by the operator
%%%%%%%%%%%%%%%%%%%%%%%%%%%%%%%%%%%%%%%%
\begin{equation}
\hat{U}(\theta)
= e^{-i \theta \hat{n}},
\label{U op 1}
\end{equation}
%%%%%%%%%%%%%%%%%%%%%%%%%%%%%%
where $\hat{n}=a^{\dagger}a$ is the number operator, and $a$, $a^{\dagger}$ are the annihilation and creation operators, respectively, given by
%%%%%%%%%%%%%%%%%%%%%%%%%%%%%%%%%%%%%%%%
$
%\begin{equation}
a = \frac{1}{\sqrt{2}}(\hat{x} + i \hat{p})
$,
$
a^{\dagger} = \frac{1}{\sqrt{2}}(\hat{x} - i \hat{p}) \; .
%\label{a,a+}  
%\end{equation}
$
%%%%%%%%%%%%%%%%%%%%%%%%%%%%%%%%
Indeed, using the operator identity 
(Ref. \cite{messiah}, p. 339)
%%%%%%%%%%%%%%%%%%%%%%%%%%%%%%%%%%%%%%%%
\begin{equation}
e^{\hat{A}} \hat{B} e^{-\hat{A}}
= \hat{B} + [\hat{A}, \hat{B}] + \frac{1}{2!}[\hat{A},[\hat{A},\hat{B}]]
+ \cdots \; ,
\label{e(A)Be(-A)}
\end{equation}
%%%%%%%%%%%%%%%%%%%%%%%%%%%%%%
we readily find that the operator (\ref{U op 1}) gives the transformation properties of $a$ and $a^{\dagger}$
%%%%%%%%%%%%%%%%%%%%%%%%%%%%%%%%%%%%%%%%
\begin{equation}
\hat{U}^{\dagger}(\theta) \hat{a} \hat{U}(\theta)
= e^{-i \theta} \hat{a}, \;\;\;\;\;
\hat{U}^{\dagger}(\theta) \hat{a}^{\dagger} \hat{U}(\theta) 
= e^{i \theta} \hat{a}^{\dagger} \;,
\label{U+aU,U+a+U}
\end{equation}
%%%%%%%%%%%%%%%%%%%%%%%%%%%%%%%%
that lead to the transformation properties of $\hat{x}$ and $\hat{p}$, 
Eq. (\ref{X=UxU,P=UpU}) [with Eq. (\ref{Xtheta,Ptheta})].
It is also shown in Appendix \ref{m.els.exp(itn)} that the matrix elements of the operator $\exp{(i\theta \hat{n})}$ (the adjoint of (\ref{U op 1})) are identical to those of Eq.  (\ref{<x'U+x>}) found earlier.

Finally, we verify that the bases
$\{|x_1;\theta_1\rangle \}$ and $\{|x_2;\theta_2\rangle\}$
(with fixed $\theta_1 \neq \theta_2$) that we have been studying above
are MUB.
Using Eq. (\ref{<x'U+x>}) and the relation 
$U(\theta_2)U^{\dagger}(\theta_1)=U^{\dagger}(\theta_1-\theta_2)$,
which follows from (\ref{U op 1}), we find
%%%%%%%%%%%%%%%%%%%%%%%%%%%%%%%%%%
\begin{equation}
|\langle x_2;\theta_2|x_1;\theta_1 \rangle|^2=
|\langle x_2|U^{\dagger}(\theta_1-\theta_2)|x_1\rangle|^2
=\frac{1}{2\pi|S(\theta_1,\theta_2)|} \; ,
\label{|<x'theta'|x theta>|}
\end{equation}
%%%%%%%%%%%%%%%%%%%%%%%%%%%%%%%%
where $S(\theta_1,\theta_2)=\sin(\theta_1 - \theta_2)$. 
The number $|\langle x_2;\theta_2|x_1;\theta_1\rangle|^2$ is thus
{\em independent} of $x_1$ and $x_2$, so that, according to definition (\ref{MUB}), the two bases are MUB.
As an example, for $\theta = \pi/2$, $|x';\theta=\frac{\pi}{2}\rangle$ is  an eigenfunction of $\hat{p}$ with eigenvalue $x'$, whose projection in the $x$ representation is $e^{ix'x}/\sqrt{2\pi}$
[see the second Eq. (\ref{lim theta to 0 and plane_wave})], its absolute value squared being consistent with Eq. (\ref{|<x'theta'|x theta>|}).

In Appendix \ref{graphical_method} we present a simple way to derive the result of Eq. (\ref{|<x'theta'|x theta>|}), which is an application of the idea of doing QM in phase space using Wigner transforms, mentioned right below Eq. (\ref{<A>}).

%%%%%%%%%%%%%%%%%%%%%%%%%%%%%%%%
\subsection{State reconstruction based on MUB}
\label{QM state recons and MUB}

Now we show that the MUB introduced above can be used to perform a quantum-mechanical state reconstruction.
We first introduce the set of operators
%%%%%%%%%%%%%%%%%%%%%%%%%%%%%%%%%%
\begin{equation}
\hat{\mathbb{Z}}(a,b)
=e^{ia\hat{x}}e^{ib\hat{p}}
= e^{-\frac{i}{2} ab} e^{i(a \hat{x} + b \hat{p})},
\;\;\;\;\; -\infty < a,b < +\infty,
\label{Z(a,b) 2}
\end{equation}
%%%%%%%%%%%%%%%%%%%%%%%%%%%%%%%%%
where we have used the BCH identity, Eq. (\ref{BCH}).
These operators form a complete and orthogonal operator basis 
\cite{schleich,ulf}.
They satisfy the orthogonality property
%%%%%%%%%%%%%%%%%%%%%%%%%%%%%%%%%%
\begin{equation}
\int {\rm Tr}\left[\hat{\mathbb{Z}}^{\dagger}(a',b') \hat{\mathbb{Z}}(a,b)\right] 
\frac{da\; db}{2 \pi}
=\delta(a'-a)\delta(b'-b).
\label{orthog. Z} 
\end{equation}
%%%%%%%%%%%%%%%%%%%%%%%%%%%%%%%%%
Thus we express the density operator as a linear combination of the operators 
$\hat{\mathbb{Z}}(a,b)$ as
%%%%%%%%%%%%%%%%%%%%%%%%%%%%
%\begin{subequations}
\begin{eqnarray} 
\hat{\rho}
=\int c(a,b) \hat{\mathbb{Z}}(a,b) \frac{da \; db}{2\pi}, \;\;\;
%\label{expansion rho}  \\
c(a,b) = {{\rm Tr} \left[\hat{\rho} \hat{\mathbb{Z}}^{\dagger}(a,b)\right]} .
%\label{c}
\label{expansion rho 1}
\end{eqnarray}
%\label{expansion rho 1}
%\end{subequations}
%%%%%%%%%%%%%%%%%%%%%%%%%%%%%%%%%
In the above equations, $a$ and $b$ play the role of Cartesian coordinates. We go over to polar coordinates, defining
$a=r \cos\theta$, $b=r \sin\theta$,
so that Eq. (\ref{expansion rho 1}) takes the form
%%%%%%%%%%%%%%%%%%%%%%%%%%%%
\begin{equation}
\hat{\rho}
= \frac{1}{2 \pi} \int_0^{\infty} dr \; r \int_0^{2 \pi} d \theta \;
{\rm Tr} \left[
\hat{\rho} \; e^{-ir(C \hat{x} + S \hat{p})}
\right]
e^{ir(C \hat{x} + S \hat{p})} \; .
\label{expansion rho 2}
\end{equation}
%%%%%%%%%%%%%%%%%%%%%%%%%%%%
[We use the abbeviations $C$ and $S$ from Eq. (\ref{Xtheta})].
Using similar arguments to those that led from 
Eq. (\ref{rho(x,y)_invFT_polar}) to (\ref{rho(x,y)_invFT_polar 2}), we rewrite the above equation as
%%%%%%%%%%%%%%%%%%%%%%%%%%%%
\begin{equation}
\hat{\rho}
= \frac{1}{2 \pi} \int_{-\infty}^{\infty} dt \; |t| \int_0^{\pi} d \theta \;
{\rm Tr} \left[
\hat{\rho} \; e^{-it(C \hat{x} + S \hat{p})}
\right]
e^{it(C \hat{x} + S \hat{p})} \; .
\label{expansion rho 2}
\end{equation}
%%%%%%%%%%%%%%%%%%%%%%%%%%%%
Since in the exponent of this last equation we have the operator $\hat{X}_{\theta}$ [see Eq. (\ref{Xtheta,Ptheta})],
the exponential can be written in its spectral representation
[see Eq. (\ref{e-value eqn for Xtheta_QM})] as
%%%%%%%%%%%%%%%%%%%%%%%%%%%%
\begin{equation} 
e^{it(C\hat{x}+S\hat{p})} 
= e^{it\hat{X}_{\theta}} 
%= \int |x',\theta\rangle e^{itx'}\langle x',\theta| dx'  
= \int e^{itx'} \hat{\mathbb{P}}_{x', \theta} dx',
\label{expitX 1c}
\end{equation}
%%%%%%%%%%%%%%%%%%%%%%%%%%%%
where the projection operator $\hat{\mathbb{P}}_{x', \theta}$ is defined in 
Eq. (\ref{rhoQM(x')}). Similarly,
%%%%%%%%%%%%%%%%%%%%%%%%%%%%
\begin{subequations}
\begin{eqnarray} 
{\rm Tr} \left[
\hat{\rho} \; e^{-it(C \hat{x} + S \hat{p})}
\right] 
&=& \int e^{-itx'} {\rm Tr}\left( \hat{\rho} 
\hat{\mathbb{P}}_{x', \theta}\right) dx'   \\
&=& \int e^{-itx'} \rho_{\theta}^{QM}(x')dx',
\label{Tr_rho_exp-itX 1c}
\end{eqnarray}
\label{Tr_rho_exp-itX 1}
\end{subequations}
%%%%%%%%%%%%%%%%%%%%%%%%%%%%
where we have used the definition of the QM probability density $\rho_{\theta}^{QM}(x')$, Eq. (\ref{rhoQM(x')}).
Then Eq. (\ref{expansion rho 2}) for $\hat{\rho}$, using 
Eqs. (\ref{expitX 1c}) and (\ref{Tr_rho_exp-itX 1c}), becomes
%%%%%%%%%%%%%%%%%%%%%%%%%%%%
\begin{subequations}
\begin{eqnarray} 
\hat{\rho}
&=& \frac{1}{2 \pi} \int_{-\infty}^{\infty} dt \; |t| 
\int_0^{\pi} d \theta \;
\int \int_{-\infty}^{\infty} dx' dx''
e^{-it(x'-x'')}\rho_{\theta}^{QM}(x')
\hat{\mathbb{P}}_{x'', \theta}    \\
&=& \frac{1}{2 \pi} `
\lim_{\epsilon \to 0^+}
\int_0^{\pi} d \theta \;
\int \int_{-\infty}^{\infty} dx' dx''
f_{\epsilon}(x'-x'')
\rho_{\theta}^{QM}(x')
\hat{\mathbb{P}}_{x'', \theta} .
\label{expansion rho 4 b}
\end{eqnarray}
\label{expansion rho 4}
\end{subequations}
%%%%%%%%%%%%%%%%%%%%%%%%%%%%
In the last line we have performed the radial integral and used the definition (\ref{f_epsilon}) that was introduced in our earlier analysis, in the course of inverting the Radon transform.
It is important to note that in the present context we have been able to express the density operator $\hat{\rho}$ {\em directly} in terms of the probability $\rho_{\theta}^{QM}(x')$, thanks to the expansion of $\hat{\rho}$, 
Eq. (\ref{expansion rho 2}), in terms of MUB, together with 
Eq. (\ref{Tr_rho_exp-itX 1c}) which relates the trace on its left-hand side with $\rho_{\theta}^{QM}(x')$.

The Wigner function for the state $\hat{\rho}$ of 
Eq. (\ref{expansion rho 4 b}) is identical to the result found above in Eq. (\ref{inverse radon q,p QM}), which we reproduce here for completeness
%%%%%%%%%%%%%%%%%%%%%%%%
\begin{equation}
W_{\hat{\rho}}(q,p)
= - \frac{1}{\pi}\int_0^{\pi} d \theta \;
{\cal P}\int_{-\infty}^{\infty} dx'
\frac{\partial \rho_{\theta}^{QM}(x')/\partial x'}
{x' - (q \cos \theta + p \sin \theta)} \; .
\label{Wfctn of rho(MUB)}
\end{equation}
%%%%%%%%%%%%%%%%%%%%%%%%%%%%%
This result can be proved as follows.
Application of Eq. (\ref{WT}) --defining the Wigner function-- to the density operator $\hat{\rho}$, Eq. (\ref{expansion rho 4 b}), gives
%%%%%%%%%%%%%%%%%%%
\begin{eqnarray}
W_{\hat{\rho}}(q,p)
&=&\frac{1}{2 \pi}
\lim_{\epsilon \to 0^+}
\int_0^{\pi} d \theta \;
\int \int \int_{-\infty}^{\infty} dy dx' dx''
f_{\epsilon}(x'-x'')
\rho_{\theta}^{QM}(x') 
\nonumber  \\
&& \;\;\;\;\; \times e^{-ipy}
\left\langle q + \frac{y}{2} 
\left\vert \hat{\mathbb{P}}_{x'', \theta} \right\vert
q - \frac{y}{2} \right\rangle.
\label{WrhoMUB 1}
\end{eqnarray}
%%%%%%%%%%%%%%%%%%%
We evaluate the matrix element of the projector
$\hat{\mathbb{P}}_{x'', \theta}$
by using its definition in Eq. (\ref{rhoQM(x')}), 
the unitary transformation, Eq. (\ref{U 1}), and its explicit expression, Eq. (\ref{<x'U+x>}), to find
%%%%%%%%%%%%%%%%%%%
\begin{equation}
\left\langle q + \frac{y}{2} 
\left\vert \hat{\mathbb{P}}_{x'', \theta} \right\vert
q - \frac{y}{2} \right\rangle
= \frac{e^{\frac{i}{\sin \theta}y(x'' - q \cos \theta)}}
{2 \pi \sin \theta}.
\label{<P> 1}
\end{equation}
%%%%%%%%%%%%%%%%%%%
Substituting this result in Eq. (\ref{WrhoMUB 1}) and performing the integration over $y$ we have
%%%%%%%%%%%%%%%%%%%
\begin{eqnarray}
W_{\hat{\rho}}(q,p)
&=& 
\lim_{\epsilon \to 0^+}\frac{1}{2 \pi}
\int_0^{\pi} d \theta \;
\int \int_{-\infty}^{\infty} dx' dx''
f_{\epsilon}(x'-x'')
\rho_{\theta}^{QM}(x') 
\nonumber \\
&& \;\;\;\;\; \times \delta(x'' - (q\cos \theta + p \sin \theta)) 
\nonumber \\
&=& 
\lim_{\epsilon \to 0^+} \frac{1}{2 \pi} 
\int_0^{\pi} d \theta \;
\int_{-\infty}^{\infty} dx' \;
f_{\epsilon}(x'- (q\cos \theta + p \sin \theta))
\rho_{\theta}^{QM}(x') \; .
\label{WrhoMUB 2}
\end{eqnarray}
%%%%%%%%%%%%%%%%%%%
The last line is $2 \pi$ times the right-hand side of 
Eq. (\ref{rho(x,y)_invFT_polar 5}), with $x$ replaced by $q$ and 
$y$ by $p$. 
We thus take over the result of Eq. (\ref{inverse radon}), making these replacements and multiplying by $2 \pi$, and find
Eq. (\ref{Wfctn of rho(MUB)}).

Finally, we calculate the matrix elements of the density operator (\ref{expansion rho 4 b}) in the coordinate representation, 
$\langle x_1 |\hat{\rho}| x_2 \rangle$, which is the counterpart in Hilbert space of Eq. (\ref{Wfctn of rho(MUB)}).
For the matrix elements of the projector $\hat{\mathbb{P}}_{x'', \theta}$
we find, just as in Eq. (\ref{<P> 1}),
%%%%%%%%%%%%%%%%%%%
\begin{equation}
\left\langle x_1
\left\vert \hat{\mathbb{P}}_{x'', \theta} \right\vert
x_2 \right\rangle
= \frac{e^{\frac{i}{\sin \theta}
(x_1 - x_2)(x'' - \frac{x_1+x_2}{2} \cos \theta)}}
{2 \pi \sin \theta},
\label{<P> 2}
\end{equation}
%%%%%%%%%%%%%%%%%%%
so that
%%%%%%%%%%%%%%%%%%%%%%%%
\begin{equation}
\langle x_1 |\hat{\rho}| x_2 \rangle
= \frac{1}{2 \pi} \lim_{\epsilon \to 0^+}\int_0^{\pi} d \theta \;
\int \int_{-\infty}^{\infty} dx' dx'' f_{\epsilon}(x'-x'')
\rho_{\theta}^{QM}(x')
\frac{e^{\frac{i}{\sin \theta}(x_1 - x_2)
(x'' - \frac{x_1 + x_2}{2}\cos \theta)}}
{2 \pi \sin \theta} \; .
\label{rho(x1,x2) 1}
\end{equation}
%%%%%%%%%%%%%%%%%%%%%%%%%%%%%
We compare this last equation with Eq. (\ref{rho(x,y)_invFT_polar 5}) and use the result of 
Eq. (\ref{inverse radon}) to obtain
%%%%%%%%%%%%%%%%%%%%%%%%
\begin{equation}
\langle x_1 |\hat{\rho}| x_2 \rangle
= - \frac{1}{\pi}\int_0^{\pi} d \theta \;
{\cal P}\int \int_{-\infty}^{\infty} \frac{dx' dx''}{x'-x''}
\frac{\partial \rho_{\theta}^{QM}(x')}{\partial x'}
\frac{e^{\frac{i}{\sin \theta}(x_1 - x_2)
(x'' - \frac{x_1 + x_2}{2}\cos \theta)}}
{2 \pi \sin \theta} \; ,
\label{rho(x1,x2)}
\end{equation}
%%%%%%%%%%%%%%%%%%%%%%%%%%%%%
which shows explicitly how $\rho_{\theta}^{QM}(x')$, which is a  probability density, and hence a measurable quantity, can be used to find the matrix elements of the density operator.

This completes our demonstration of the consistency of the two approaches to the problem of quantum-mechanical state reconstruction that we have considered in this paper, for systems described in a continuous Hilbert space.
On the one hand, the approach presented in the previous section based on tomography in phase space and the Wigner function and, on the other, the one given in the present section based on the expansion of the density operator in terms of operators defined via MUB.
The difference in the strategies of these two approaches involves, essentially, two ways of handling the complete orthonormal operators 
$e^{ia \hat{x}}e^{ib \hat{p}}$:
the Wigner function approach that led us to the results in 
Sec. \ref{quantum reconstruction} can be regarded as using these operators
to construct the Fourier transform of the density operator, as in 
Eq. (\ref{WT=invFT and char_fctn}).
If, on the other hand, we consider their spectral representation, 
Eq. (\ref{expitX 1c}), we are led to the MUB approach of the present 
section.

%%%%%%%%%%%%%%%%%%%%%%%%%
\section{Mutually unbiased bases and state reconstruction in a finite-dimensional Hilbert space}
\label{finite mub}

Considerable work has been devoted to the study of MUB in a finite, $d$-dimensional Hilbert space \cite{amir,tal,combescure,ivanovich}.
In this paper we restrict our study to the case in which the dimensionality $d$ is a {\em prime number}:
for this case the number of MUB is exactly $d+1$
\cite{wootters2,gibbons_et_al_2004,tal}.
The finite-dimensional theory is intriguingly connected with sophisticated mathematical notions 
\cite{combescure,vourdas1-klimov1-planat1-klimov2}
%\cite{klimov1,vourdas1,planat1,combescure,klimov2}, 
that we do not consider here.

In the finite, $d$-dimensional Hilbert space problem,
where a Radon-like transform is not available, we shall follow a procedure which is analogous to that presented in the last section
for a continuous, infinite-dimensional Hilbert space.

We first consider the $d$-dimensional Hilbert space to be spanned by $d$ distinct states 
$|n\rangle$, with $n=0,1, \cdots ,(d-1)$, which are subject to the periodic condition
$|n+d\rangle=|n\rangle$.
These states are designated as the ``computational basis" of the space.  
We shall follow Schwinger \cite{schwinger} and introduce the unitary operators $\hat{X}$ and $\hat{Z}$, which play a role analogous to that of the position operator $\hat{x}$ and the momentum operator $\hat{p}$ of the continuous case. 
The Schwinger operators are defined by their action on the states of the computational basis by the equations
%%%%%%%%%%%%%%%%%%%%%%%%%%%%%%%
\begin{subequations}
\begin{eqnarray}
\hat{Z}|n\rangle
&=&\omega^n|n\rangle, \;\;\;\; \omega=e^{2 \pi i/d},
\label{Z}  \\
\hat{X}|n\rangle &=& |n+1\rangle .
\label{X}
\end{eqnarray}
\label{Z,X}
\end{subequations}
%%%%%%%%%%%%%%%%%%%%%%%%%%%%%%
These definitions lead to the commutation relation
%%%%%%%%%%%%%%%%%%%%%%%%%%%%%%
\begin{equation}
\hat{Z}\hat{X}=\omega \hat{X}\hat{Z} .
\label{comm Z,X}
\end{equation}
%%%%%%%%%%%%%%%%%%%%%%%%%%%%%%
The two operators $\hat{Z}$ and $\hat{X}$ form a complete algebraic set, in that only a multiple of the identity commutes with both \cite{schwinger}. 
As a consequence, any operator defined in our $d$-dimensional Hilbert space can be written as a function of $\hat{Z}$ and $\hat{X}$.

The $d^2$-dimensional matrix space is spanned by the complete orthonormal $d^2$ operators $\hat{X}^m \hat{Z}^l$, with $m,l=0,1,..(d-1)$, so that any 
$d \times d$ matrix can be written as a linear combination of these $d^2$ operators.
A familiar example is a $2$-dimensional Hilbert space, where any 
$2\times 2$ matrix can be written as a linear combination of the three Pauli matrices plus the unit matrix, which can also be written as $\sigma_x$, $\sigma_z$, $\sigma_x \sigma_z$ and $I$.

The operators $\hat{X}^m \hat{Z}^l$ are orthonormal under the trace operation,
%%%%%%%%%%%%%%%%%%%%%%%%%%%%%%
\begin{equation} 
{\rm Tr}\left[\hat{X}^m\hat{Z}^l\left(\hat{X}^{m'}\hat{Z}^{l'}\right)^{\dagger}\right]
= d \; \delta_{m,m'}\delta_{l,l'},
\label{orth}
\end{equation}
%%%%%%%%%%%%%%%%%%%%%%%%%%%%%%
a relation which can be proved directly using the defining Eqs. (\ref{Z,X}).
Completeness follows from the set consisting of $d^2$ linearly independent operators. 

We shall need to do arithmetic operations on the numbers  
$n=0,1,\cdots ,(d-1)$ that label our states, assuming the periodic condition $d=0[{\rm mod} \; d]$.
When $d$ is a prime number, the operations of multiplication and division, modulo $d$, can be defined consistently \cite{schroder}.
As a simple example, we find, for $d=3$, that $1/2[{\rm mod} \; 3]=2$, since $2\cdot2=4=1[{\rm mod} \; 3]$.

As a result, we may replace uniquely, up to a power of $\omega$, all the operators of the form $\hat{X}^m\hat{Z}^l$, with $m\ne 0$, by $(\hat{X}\hat{Z}^b)^m$, with $b=lm^{-1}=0,1,...(d-1)$.
We can readily prove that 
%%%%%%%%%%%%%%%%%%%%%%%%%%%%%%
\begin{equation} 
\hat{X}^m\hat{Z}^l
= \omega ^{-\frac{m(m-1)}{2}b} (\hat{X} \hat{Z}^{b})^m \; ,
\label{X2Z vs XZ1/2 }
\end{equation}
%%%%%%%%%%%%%%%%%%%%%%%%%%%%%%
with $l=mb[{\rm mod}\; d]$.
>From Eq. (\ref{orth}) we find, for the new quantities $(XZ^b)^m$, the orthogonality relation
%%%%%%%%%%%%%%%%%%%%%%%%%%%%%%
\begin{equation}
{\rm
Tr}\left[(\hat{X}\hat{Z}^b)^{m}\left((\hat{X}\hat{Z}^{b'})^{m'}\right)^\dagger\right]
= d \; \delta_{b,b'}\delta_{m,m'},\;\;\; m,m'\ne 0.
\label{orth 1}
\end{equation}
%%%%%%%%%%%%%%%%%%%%%%%%%%%%%%
The operators $(\hat{X}\hat{Z}^b)^m$ are $d(d-1)$ in number; these, together with the $d$ operators $\hat{Z}^l$, with $l=0,1,...(d-1)$
(this last set evidently contains the identity: $\hat{Z}^0=\hat{Z}^d=I$),
form a complete orthonormal set of $d^2$ operators which is equivalent to the set $\hat{X}^m\hat{Z}^l$ considered above. 
Thus we may express an arbitrary density operator as a linear combination of these operators as
%%%%%%%%%%%%%%%%%%%%%%%
\begin{eqnarray}
\hat{\rho}
&=&\frac{1}{d}\left\{\sum_{b=0}^{d-1} \sum_{m=1}^{d-1}
{\rm Tr} \Big[ \hat{\rho} ((\hat{X} \hat{Z}^b)^m)^{\dagger} \Big]
(\hat{X} \hat{Z}^b)^m 
%\right.
%\nonumber \\
%&& \hspace{15mm} \left. 
+ \sum_{l=0}^{d-1}{\rm Tr}
\Big[\hat{\rho} (\hat{Z}^l)^{\dagger} \Big]
\hat{Z}^l \right\}.
\label{rho-lc-discr-mub}
\end{eqnarray}
%%%%%%%%%%%%%%%%%%%%%%%

For a given $b$, the operator $\hat{X}\hat{Z}^b$ possesses $d$ eigenvectors, which we denote by
$|c;b\rangle$, $c=0,1, \cdots, d-1$.
In terms of the computational basis these eigenvectors are given by \cite{tal}
%%%%%%%%%%%%%%%%%%%%%%%
\begin{equation}
|c;b\rangle
=\frac1{\sqrt{d}}\sum_{n=0}^{d-1}\omega^{\frac b2 n(n-1)-cn}|n\rangle ,
\;\;\;\;\; 
\hat{X}\hat{Z}^b|c;b\rangle=\omega^c|c;b\rangle.
\label{mubstate}
\end{equation}
%%%%%%%%%%%%%%%%%%%%%%%
This equation defines $d$ distinct bases ($b=0,1,\cdots,d-1$) which, when supplemented with the computational basis,
which is an eigenbasis of the operator $\hat{Z}$ [see Eq. (\ref{Z})],
forms a set of $d+1$ MUB bases, i.e., 
%%%%%%%%%%%%%%%%%%%%%%%
\begin{subequations}
\begin{eqnarray}
\langle c;b|c'; b\rangle 
&=& \delta_{c,c'}, \;\;\;
|\langle c;b|c';b'\rangle|^2 = \frac1{d},\;\;b\ne b', \\
\langle n |n' \rangle &=& \delta _{n,n'},  \;\;\;\;\;\;
|\langle n|c;b\rangle|^2 = \frac1{d} \; .
\end{eqnarray}
\end{subequations}
%%%%%%%%%%%%%%%%%%%%%%%
These equations can be proved straightforwardly by direct evaluation. 

We rewrite Eq. (\ref{rho-lc-discr-mub}) by adding and subtracting the $m=0$ terms as
%%%%%%%%%%%%%%%%%%%%%%%
\begin{eqnarray}
\hat{\rho}
&=&\frac{1}{d}\left\{\sum_{b=0}^{d-1} \sum_{m=0}^{d-1}
{\rm Tr} \Big[ \hat{\rho} ((\hat{X} \hat{Z}^b)^m)^{\dagger} \Big]
(\hat{X} \hat{Z}^b)^m 
%\right.
%\nonumber \\
%&& \hspace{15mm} \left. 
-d \; \Bbb{I}
+ \sum_{l=0}^{d-1}{\rm Tr}
\Big[\hat{\rho} (\hat{Z}^l)^{\dagger} \Big]
\hat{Z}^l \right\}.
\label{rho-lc-discr-mub 1}
\end{eqnarray}
%%%%%%%%%%%%%%%%%%%%%%%
The spectral representation of the operator $\hat{X}\hat{Z}^b$ is given by
%%%%%%%%%%%%%%%%%%%%%%%
\begin{equation}
\hat{X}\hat{Z}^b=\sum_{c=0}^{d-1}|c;b\rangle\omega^c\langle c;b|.
\end{equation}
%%%%%%%%%%%%%%%%%%%%%%%
Note that the eigenvalues $\omega^c$ are non-degenerate.
We obtain
%%%%%%%%%%%%%%%%%%%%%%%
\begin{equation}
{\rm Tr}[\hat{\rho}((\hat{X}\hat{Z}^b)^m)^{\dagger}]
=\sum_{c=0}^{d-1}\langle c;b|\hat{\rho}|c;b\rangle \omega^{-cm}.
\end{equation}
%%%%%%%%%%%%%%%%%%%%%%%
Hence the density operator of Eq. (\ref{rho-lc-discr-mub 1}) takes the form
%%%%%%%%%%%%%%%%%%%%%%%
\begin{equation}
\hat{\rho}=\sum_{b,c =0}^{d-1}|c;b \rangle \langle c;b|\hat{\rho}|c;b\rangle 
\langle c;b|
+\sum_{n=0}^{d-1}|n\rangle
\langle n|\hat{\rho}|n \rangle \langle n|
-\Bbb{I}.
\label{rho(discret-mub)}
\end{equation}
%%%%%%%%%%%%%%%%%%%%%%
The matrix elements of $\hat{\rho}$ in the computational basis are then given by
%%%%%%%%%%%%%%%%%%%%%%
\begin{equation}
\langle n'|\hat{\rho}|n'' \rangle
=\sum_{b,c =0}^{d-1}\langle n'|c;b \rangle \langle c;b|\hat{\rho}|c;b\rangle 
\langle c;b|n''\rangle
+ \langle n'|\hat{\rho}|n' \rangle \delta_{n',n''}
- \delta_{n',n''} .
\end{equation}
%%%%%%%%%%%%%%%%%%%%%%
The density operator $\hat{\rho}$ is given in terms of probabilities, 
Eq. (\ref{rho(discret-mub)}), which are observable quantities;
e.g., $\langle c;b|\hat{\rho}|c;b \rangle$ is the probability to find the state $|c;b\rangle$ when the system is described by the density operator $\hat{\rho}$. 
We thus find that $\hat{\rho}$ is reconstructed by using $d+1$ measurements
\cite{ivanovich}. 
Each of these measurements yields $d-1$ independent probability outcomes (since the probabilities add up to 1). 
This gives $(d+1)(d-1)=d^2-1$ quantities, which is precisely the number of independent parameters of a $d$-dimensional density matrix. 

Finally, we wish to call the reader's attention to the analogy between several quantities used in the present section and those introduced in the previous one, where a continuous, infinite-dimensional Hilbert space was used.
This correspondence is indicated in Table \ref{analogy-discr-cont}.
%%%%%%%%%%%%%%%%%%%%%%%%%
%\begin{eqnarray}
%\begin{array}{ccc}
% {\rm \underline{Discrete \; case}}   &  \hspace{1cm}  &  
%{\rm \underline{Continuous \; case}}   \\
%\hat{X}^m \hat{Z}^l &  \hspace{1cm} &  e^{ia\hat{x}}  e^{ib\hat{p}}          
%\\
%\hat{X}^m \hat{Z}^{mb}=\omega^{-mb}(\hat{X}\hat{Z}^b)^m 
%&   \hspace{1cm}     &        e^{ia\hat{x}}  e^{ib\hat{p}} 
%=e^{-\frac{i}{2}r\cdot rCS}\left[e^{i(C\hat{x} + S\hat{p})}\right]^r
%\\
%| c; b \rangle   &       &  |x'; \theta  \rangle    \\
%\hat{X}\hat{Z}^b | c; b \rangle 
%= \omega^c | c; b \rangle 
%&         & e^{i(C\hat{x} + S\hat{p})} |x'; \theta  \rangle
%=e^{ix'} |x'; \theta  \rangle
%\\
%(\hat{X}\hat{Z}^b)^m | c; b \rangle 
%= \omega^{mc} | c; b \rangle 
%&       & e^{ir(C\hat{x} + S\hat{p})} |x'; \theta  \rangle
%=e^{irx'} |x'; \theta  \rangle
%\end{array}
%\end{eqnarray}
%%%%%%%%%%%%%%%%%%%%%%%%%
 
%%%%%%%%%%%%%%%%%%%%%%%%%
\begin{table}[tbp]
\caption{\footnotesize Analogy between quantities for a discrete and a continuous Hilbert space}
\begin{tabular}{||cc||}
\hline\hline
Discrete case &  \multicolumn{1}{|c||}{Continuous case}  
\\ \hline\hline
$\hat{X}^m \hat{Z}^l$ & \multicolumn{1}{|c||}{$e^{ia\hat{x}}  e^{ib\hat{p}}$} 
\\ \hline
$\hat{X}^m \hat{Z}^{l}=\omega^{-\frac{m(m-1)}{2}b}(\hat{X}\hat{Z}^b)^m$ & \multicolumn{1}{|c||}{$e^{ia\hat{x}}  e^{ib\hat{p}} 
=e^{-i\frac{r^2}{2}\cdot CS}\left[e^{i(C\hat{x} + S\hat{p})}\right]^r$} 
\\ \hline
$| c; b \rangle$ & \multicolumn{1}{|c||}{$|x'; \theta  \rangle$} 
\\ \hline
$\hat{X}\hat{Z}^b | c; b \rangle 
= \omega^c | c; b \rangle$ & \multicolumn{1}{|c||}{$e^{i(C\hat{x} + S\hat{p})} |x'; \theta  \rangle
=e^{ix'} |x'; \theta  \rangle$} 
\\ \hline
$(\hat{X}\hat{Z}^b)^m | c; b \rangle 
= \omega^{mc} | c; b \rangle$ & \multicolumn{1}{|c||}{$\left[e^{i(C\hat{x} + S\hat{p})}\right]^r |x'; \theta  \rangle
=e^{irx'} |x'; \theta  \rangle$}  
\\ \hline\hline
\end{tabular}
\label{analogy-discr-cont}
\end{table}
%%%%%%%%%%%%%%%%%%%%%%%%%

%%%%%%%%%%%%%%%%%%%%%%%%%
%\begin{eqnarray}
%{\rm \underline{Discrete \; case}}   &  \hspace{1cm}  &  
%{\rm \underline{Continuous \; case}}   
%\nonumber \\
%\hat{X}^m \hat{Z}^l &  \hspace{1cm} &  e^{ia\hat{x}}  e^{ib\hat{p}}          
%\\
%\hat{X}^m \hat{Z}^{l}=\omega^{-\frac{m(m-1)}{2}b}(\hat{X}\hat{Z}^b)^m 
%&   \hspace{1cm}     &        e^{ia\hat{x}}  e^{ib\hat{p}} 
%=e^{-i\frac{r^2}{2}\cdot CS}\left[e^{i(C\hat{x} + S\hat{p})}\right]^r
%\;\;\;    \\
%| c; b \rangle   &       &  |x'; \theta  \rangle    \\
%\hat{X}\hat{Z}^b | c; b \rangle 
%= \omega^c | c; b \rangle 
%&         & e^{i(C\hat{x} + S\hat{p})} |x'; \theta  \rangle
%=e^{ix'} |x'; \theta  \rangle
%\\
%(\hat{X}\hat{Z}^b)^m | c; b \rangle 
%= \omega^{mc} | c; b \rangle 
%&       & \left[e^{i(C\hat{x} + S\hat{p})}\right]^r |x'; \theta  \rangle
%=e^{irx'} |x'; \theta  \rangle
%\end{eqnarray}
%%%%%%%%%%%%%%%%%%%%%%%%%
 
%%%%%%%%%%%%%%%%%%%%%%%%%
\section{Conclusions and Remarks}
\label{concl}

We have reviewed the approach to the quantum-state reconstruction problem based on the Wigner function and the Radon transform, pointing out its close analogy with classical tomography.
We put emphasis on the role played by the Wigner function, which was shown to be analogous to that of the probability density in phase space for the classical problem.

The analysis underscores the intriguing fact that to reconstruct a quantum state we require the probabilities of all the phase-space plane, and not merely the probabilities along the position and momentum axes as might be implied by a positive reply to Pauli's query posed in the Introduction.

Then we reviewed an alternative route for the state reconstruction which is based on
the notion of mutually unbiased bases and does not make use of the Radon transform.
We described its connection with the method based on the Wigner function. 

In addition, we showed that the concept of mutually unbiased bases can be applied to the problem of state reconstruction for a finite-dimensional Hilbert space, which is quite relevant for all applications to quantum computing.
Finally, a parallel with the case of a continuous, infinite-dimensional Hilbert space is drawn.

\acknowledgments

F. C. K. acknowledges financial support from NSERCC.
P. A. M. and M. R. express their gratitude to the Physics Department of the University of Alberta, Canada, where part of this work was carried out, for its hospitality.
P. A. M. acknowledges financial support from CONACyT, M\'exico, through grant No. 79501, as well as from the Sistema Nacional de Investigdores, M\'exico.
Informative discussions with Professors J. Zak, A. Mann and O. Kenneth
are gratefully acknowledged.

%%%%%%%%%%%%%%%%%%%%%%%%%
\appendix

\section{Inverting the Radon Transform: 
proof of Eq. (\ref{inverse radon})}
\label{invert_radon}

Multiplying both sides of Eq. (\ref{radon}) by $e^{-ikx'}$ and integrating over $x'$ we find
%%%%%%%%%%%%%%%%%%%%%%%%
\begin{equation}
\int_{\infty}^{\infty}
e^{-ikx'}\rho_{\theta}(x') dx'
=  \int e^{-ik(Cx + Sy)}\rho(x, y) dx dy .
\label{FTrho(x'),FTrho(x,y) 1}
\end{equation}
%%%%%%%%%%%%%%%%%%%%%%%%%%%%%
We identify the two sides of this equation with the Fourier transform
$\tilde{\rho}_{\theta}(k)$ of $\rho_{\theta}(x')$, and the Fourier transform 
$\tilde{\rho}(k_x,k_y)$ of $\rho(x, y)$, respectively, so that
%%%%%%%%%%%%%%%%%%%%%%%%
\begin{equation}
\tilde{\rho}_{\theta}(k)
= \tilde{\rho}(k_x= Ck, k_y= Sk) ,\;\;\;
k \in (-\infty, \infty).
\label{FTrho(x'),FTrho(x,y) 2}
\end{equation}
%%%%%%%%%%%%%%%%%%%%%%%%%%%%%
We recover $\rho(x, y)$ as the inverse Fourier transform of $\tilde{\rho}(k_x,k_y)$, 
%%%%%%%%%%%%%%%%%%%%%%%%
\begin{equation}
\rho(x,y)
= \frac{1}{(2\pi)^2}\int_{\infty}^{\infty} dk_x dk_y \;
e^{i(k_x x + k_y y)} \tilde{\rho}(k_x,k_y) \; ,
\label{rho(x,y)_invFT}
\end{equation}
%%%%%%%%%%%%%%%%%%%%%%%%%%%%%
where $k_x$ and $k_y$ are the Cartesian components of a wave number vector 
${\bf k}$; in polar coordinates we have
%%%%%%%%%%%%%%%%%%%%%%%%%%%
\begin{subequations}
\begin{eqnarray}
k_x &=& K \cos \phi, \;\;\;   k_y = K \sin \phi,  
\label{kx,ky(K,phi)}  \\
K &=& |{\bf k}| > 0 .
\label{K}
\end{eqnarray}
\label{kx,ky,K}
\end{subequations}
%%%%%%%%%%%%%%%%%%%%%%%
The density $\rho(x,y)$ becomes
%%%%%%%%%%%%%%%%%%%%%%%%%%%%%
\begin{equation}
\rho(x,y)
= \frac{1}{(2\pi)^2}
\int_{0}^{\infty} dK \; K  \int_0^{2\pi} d \phi \;
e^{iK(x \cos \phi + y \sin \phi)} \tilde{\rho}(K \cos \phi, K \sin \phi).
\label{rho(x,y)_invFT_polar}
\end{equation}
%%%%%%%%%%%%%%%%%%%%%%%%%%%%%
While the variable $k$ in Eq. (\ref{FTrho(x'),FTrho(x,y) 2}) is defined in the interval $(-\infty, \infty)$, the radial variable $K$ in 
Eq. (\ref{rho(x,y)_invFT_polar}) is defined to be non-negative and in the interval $(0, \infty)$.
The range of integration of the variable $K$ can be extended to the full real axis by 
first splitting the interval of integration of $\phi$ into the intervals
$(0,\pi)$ and $(\pi,2\pi)$ and then making the change of variables
$\phi = \phi' + \pi$, $K=-K'$ in the integral over the second interval,
%%%%%%%%%%%%%%%%%%%%%%%%%%%%%
to obtain
%%%%%%%%%%%%%%%%%%%%%%%%
\begin{equation}
\rho(x,y)
= \frac{1}{(2\pi)^2}\int_{-\infty}^{\infty} dk \; |k|  \int_0^{\pi} d \theta \;
e^{ik(x \cos \theta + y \sin \theta)} 
\tilde{\rho}(k \cos \theta, k \sin \theta) .
\label{rho(x,y)_invFT_polar 2}
\end{equation}
%%%%%%%%%%%%%%%%%%%%%%%%%%%%%
We identify the last factor with the quantity 
$\tilde{\rho}_{\theta}(k)$, Eq. (\ref{FTrho(x'),FTrho(x,y) 2}),
and substitute $\tilde{\rho}_{\theta}(k)$ from the left-hand side of 
Eq. (\ref{FTrho(x'),FTrho(x,y) 1}) to write
%%%%%%%%%%%%%%%%%%%%%%%%
\begin{equation}
\rho(x,y)
= \frac{1}{(2\pi)^2} \int_{-\infty}^{\infty} dk \; |k|  \int_0^{\pi} d \theta \;
e^{ik(x \cos \theta + y \sin \theta)}
\int_{-\infty}^{\infty} dx' e^{-ikx'}\rho_{\theta}(x').
\label{rho(x,y)_invFT_polar 4}
\end{equation}
%%%%%%%%%%%%%%%%%%%%%%%%%%%%%

Defining the integral
%%%%%%%%%%%%%%%%%%%%%%%%
\begin{equation}
f_{\epsilon}(\xi) \equiv \int_{-\infty}^{\infty}
|k| e^{-ik \xi - |k| \epsilon } dk, \;\;\; \epsilon >0,
\label{f_epsilon}
\end{equation}
%%%%%%%%%%%%%%%%%%%%%%%%%%%%%
and identifying
%%%%%%%%%%%%%%%%%%%%%%%%
$
%\begin{equation}
\xi = x' - (x \cos \theta + y \sin \theta),
%\label{xi=>}
%\end{equation}
$
%%%%%%%%%%%%%%%%%%%%%%%%%%%%%
we write Eq. (\ref{rho(x,y)_invFT_polar 4}) as
%%%%%%%%%%%%%%%%%%%%%%%%
\begin{equation}
\rho(x,y)
= \frac{1}{(2\pi)^2} 
\lim_{\epsilon \to 0^+}
\int_0^{\pi} d \theta \;
\int_{-\infty}^{\infty} dx' \;
f_{\epsilon}(x' - (x \cos \theta + y \sin \theta)) \;
\rho_{\theta}(x').
\label{rho(x,y)_invFT_polar 5}
\end{equation}
%%%%%%%%%%%%%%%%%%%%%%%%%%%%%
Thus our task is to study the function $f_{\epsilon}(\xi)$, which we write as
%%%%%%%%%%%%%%%%%%%%%%%%
\begin{eqnarray}
f_{\epsilon}(\xi) 
&=& \int_{-\infty}^{0}
(-k) e^{-ik (\xi + i \epsilon)} dk
+ \int_{0}^{\infty}
k e^{-ik (\xi - i \epsilon)} dk
\nonumber \\
&=&
\frac{\partial}{\partial \xi}
\left(\frac{1}{\xi + i \epsilon}
+\frac{1}{\xi - i \epsilon}\right)
\equiv \frac{\partial g_{\epsilon}(\xi) }{\partial\xi},
\label{f_epsilon 1}
\end{eqnarray}
%%%%%%%%%%%%%%%%%%%%%%%%%%%%%
where
%%%%%%%%%%%%%%%%%%%%%%%%
\begin{equation}
 g_{\epsilon}(\xi) = 2 \frac{\xi}{\xi^2 + \epsilon^2} \; .
\label{psi_epsilon}
\end{equation}
%%%%%%%%%%%%%%%%%%%%%%%%%%%%%
Using the abbreviation $\alpha = x \cos \theta + y \sin \theta$, we write the last integral in Eq. (\ref{rho(x,y)_invFT_polar 5}) as
%%%%%%%%%%%%%%%%%%%%%%%%
%\begin{subequations}
\begin{eqnarray}
I_{\epsilon} &\equiv& \int_{-\infty}^{\infty}
\rho_{\theta}(x')f_{\epsilon}(x' -\alpha) dx'  
%&=&
%\int_{-\infty}^{\infty}
%\rho_{\theta}(x')\frac{\partial g_{\epsilon}(x' -\alpha)}{\partial x'} dx' \\
=
-\int_{-\infty}^{\infty}
\frac{\partial \rho_{\theta}(x')}{\partial x'}g_{\epsilon}(x' -\alpha) dx'
\label{I-epsilon}
\end{eqnarray}
%\end{subequations}
%%%%%%%%%%%%%%%%%%%%%%%%%%%%%
where we have used the definition (\ref{f_epsilon 1}) and we have integrated by parts, assuming the integrated term to vanish for sufficiently large values of the argument.

\begin{figure}[h]
\epsfxsize=0.3\textwidth
\epsfysize=0.3\textwidth  
\centerline{\epsffile{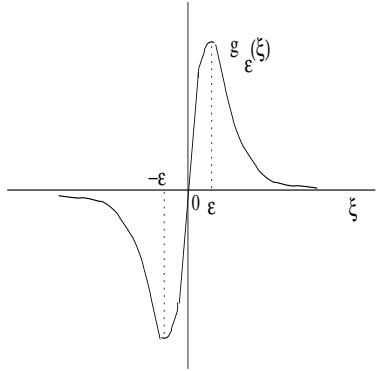}}
\caption{
\footnotesize{Schematic plot of the function $g_{\epsilon}(\xi)$ defined in Eq. (\ref{psi_epsilon})}.
%As $\epsilon \to 0$, the integral of Eq. (\ref{I-epsilon}) tends to the 
%principal-value integral of Eq. (\ref{I-epsilon->0}).
}
\label{g}
\end{figure}

The function $g_{\epsilon}(\xi)$ is shown schematically in Fig. {\ref{g}}.
%%%%%%%%%%%%%%%%%%%%%%%%%%%%%%%%%%%%%%
%%%%%%%%%%%%%%%%%%%%%%%%%%%%%%%%%%%%%%
As $\epsilon \to 0$, the integral of Eq. (\ref{I-epsilon}) tends to the principal-value integral
%%%%%%%%%%%%%%%%%%%%%%%%
\begin{equation}
\lim_{\epsilon \to 0} I_{\epsilon}
= - 2{\cal P} \int_{-\infty}^{\infty}
\frac{\partial \rho_{\theta}(x') / \partial x'}{x' - \alpha} dx' .
\label{I-epsilon->0}
\end{equation}
%%%%%%%%%%%%%%%%%%%%%%%%%%%%%
Substituting this result in Eq. (\ref{rho(x,y)_invFT_polar 5}),
%%%%%%%%%%%%%%%%%%%%%%%%
%\begin{equation}
%\rho(x,y)
%= -\frac{1}{2\pi^2} 
%\int_0^{\pi} d \theta \;
%{\cal P} \int_{-\infty}^{\infty}
%\frac{\partial \rho_{\theta}(x') / \partial x'}
%{x' - ( x \cos \theta + y \sin \theta)} dx' ,
%\label{rho(x,y)_invFT_polar 6}
%\end{equation}
%%%%%%%%%%%%%%%%%%%%%%%%%%%%%
we then find Eq. (\ref{inverse radon}) in the text.
%%%%%%%%%%%%%%%%%%%%%%%%%%%%%

%%%%%%%%%%%%%%%%%%%%%%%%%%%%%%%%
\section{Proof of Eq. (\ref{WT_P_x'_theta})}
\label{WT of Projector(theta)}

We first remark that it is easy to prove the operator identity
%%%%%%%%%%%%%%%%%
\begin{equation}
\mathbb{P}_{x'}^{\theta}
= \delta (x' - \hat{X}_{\theta}) \; .
\label{P_theta,x'=delta(x'-X)}
\end{equation}
%%%%%%%%%%%%%%%%%%%
Therefore, we compute the required Wigner transform of the projection operator (\ref{P_theta,x'=delta(x'-X)}) as
%%%%%%%%%%%%%%%%%
\begin{subequations}
\begin{eqnarray}
W_{\mathbb{P}_{x'}^{\theta}}(q,p)
&=& W_{\delta (x' - \hat{X}_{\theta})}(q,p) \\
&=& \int\left\langle q + \frac{y}{2} \Big{|}
\delta[x' - (C \hat{q} + S \hat{p})] \Big{|} q - \frac{y}{2} \right\rangle    
e^{-ipy}dy
\end{eqnarray}
\end{subequations}
%%%%%%%%%%%%%%%%%%%
It is convenient to work with the Fourier transform of this last expression with respect to the variable $x'$; i.e.,
%%%%%%%%%%%%%%%%%%%%%%%%%%%%%%%%%%%%%%%%
\begin{subequations}
\begin{eqnarray}
&&\int e^{ikx'} W_{\mathbb{P}_{x'}^{\theta}}(q,p) dx'
\\
&& \;\;\;\;\; = \int\left\langle q + \frac{y}{2} \Big{|}
e^{ik (C \hat{q} + S \hat{p})}
\Big{|} q - \frac{y}{2} \right\rangle 
e^{-ipy}dy  
\\
&& \;\;\;\;\; = e^{\frac{i}{2}k^2 SC}
\int\left\langle q + \frac{y}{2} \Big{|}
e^{ik C \hat{q}} e^{ik S \hat{p}}
\Big{|} q - \frac{y}{2} \right\rangle 
e^{-ipy}dy  \; ,
\label{FT-WPx'c}
\end{eqnarray}
\label{FT-WPx'}
\end{subequations}
%%%%%%%%%%%%%%%%%%%%%%%%%%%%%%%%%%%%%%%%
where use was made of the Baker-Campbell-Hausdorff (BCH) identity 
(Ref. \cite{messiah}, p. 442)
%%%%%%%%%%%%%%%%%%%%%%%%%%%%%%%
\begin{equation}
e^{\hat{A} + \hat{B}}
= e^{\hat{A}} \; e^{\hat{B}} \; e^{-\frac12 [\hat{A}, \hat{B}]} \; ,
\label{BCH}
\end{equation}
%%%%%%%%%%%%%%%%%%%%%%%%%%%%%%%
valid for any two Hermitean operators  $\hat{A}$, $\hat{B}$, whose commutator commutes with each of them, i.e.,
%%%%%%%%%%%%%%%%%%%%%%%%%%%
$
[\hat{A},[\hat{A}, \hat{B}]]
=[\hat{B},[\hat{A}, \hat{B}]]=0.
$
%%%%%%%%%%%%%%%%%%%%%%%%%%%
Introducing inside the matrix element of Eq. (\ref{FT-WPx'c}) a complete set of eigenstates of position and of momentum right after the first and second exponentials, respectively, we find
%%%%%%%%%%%%%%%%%%%%%%%%%%%%%%%
\begin{equation}
\int e^{ikx'} W_{\mathbb{P}_{x'}^{\theta}}(q,p) dx'
= e^{ik (C q + S p)}.
\end{equation}
%%%%%%%%%%%%%%%%%%%%%%%%%%%%%%%
The inverse Fourier transform of this last expression gives the result 
of Eq. (\ref{WT_P_x'_theta}).

%%%%%%%%%%%%%%%%%%%%%%%%%%%%%%%%%%%%
\section{Proof of Eq. (\ref{psi_{x,theta}(x') 2})}
\label{finding psi_x,theta(x')}

To prove Eq. (\ref{psi_{x,theta}(x') 2}) we impose the requirements of
Eq. (\ref{norm,m.els.X,P,theta to 0,plane_wave}) on the solution,
Eq. (\ref{psi_{x,theta}(x') 1}).

1) The ortho-normalization condition, Eq. (\ref{orthonormal}), imposed on the 
wave function $\psi_{x', \theta}(x)$ of Eq. (\ref{psi_{x,theta}(x') 1}) gives, for the function $F(x', \theta)$,
%%%%%%%%%%%%%%%%%%%%
\begin{equation}
F(x', \theta)
= \frac{e^{i \phi_{\theta}(x')}}{\sqrt{2\pi |\sin \theta|}};
\label{F 1}
\end{equation}
%%%%%%%%%%%%%%%%%%%%%%%%%%
$\phi_{\theta}(x')$ is an arbitrary phase, dependent on $x'$ and $\theta$.
The wave function $\psi_{x', \theta}(x)$
%Eq. (\ref{psi_{x,theta}(x') 1}), 
becomes
%%%%%%%%%%%%%%%%%%%%%%%%%%%%%%%%%
\begin{equation}
\psi_{x', \theta}(x)
=\frac{1}{\sqrt{2\pi |\sin \theta|}} 
e^{-\frac{i}{2 \sin \theta}
\left[ (x^2\cos \theta - 2x x')+i\phi_{\theta}(x')\right]}.
\label{psi_(x,theta)(x') 3}
\end{equation}
%%%%%%%%%%%%%%%%%%%%%%%%%%%%%%%%

2) The first requirement in Eq. (\ref{m.els. X and P}) for the matrix elements of $\hat{X}_{\theta}$
with the wave function of Eq. (\ref{psi_(x,theta)(x') 3}) is automatically fulfilled, since
i) our starting point has been the eigenvalue equation,
Eq. (\ref{e-value eqn for Xtheta_QM}), and 
ii) the wave function of Eq. (\ref{psi_(x,theta)(x') 3}) satisfies the orthonormalization condition, Eq. (\ref{orthonormal}).

3) We first compute the matrix element of $\hat{P}_{\theta}$ which appears on the left-hand side of the second requirement in 
Eq. (\ref{m.els. X and P}) . 
Using the definition of $\hat{P}_{\theta}$ given in Eq. (\ref{Xtheta,Ptheta}) and Eq. (\ref{psi_(x,theta)(x') 3}) we have, in the coordinate representation
%%%%%%%%%%%%%%%%%%%%%%%%%%%%%%%
\begin{eqnarray}
\langle x_1, \theta |\hat{P}_{\theta}| x_2, \theta \rangle
&=&\int \psi_{x_1, \theta}^{*}(x)  
\left[-\sin \theta \; x -i \cos \theta \frac{\partial}{\partial x} \right]
\psi_{x_2, \theta}(x) dx,
\nonumber \\
&=& \int \psi_{x_1, \theta}^{*}(x)  
\left[-\frac{1}{\sin \theta} x + \frac{\cos \theta}{\sin \theta}x_2\right]
\psi_{x_2, \theta}(x) dx,
\nonumber \\
&=&  -i \; e^{i\left[ \phi_{\theta}(x_2)-\phi_{\theta}(x_1)\right] } \delta ' (x_1-x_2)
+ \frac{\cos \theta}{\sin \theta} \; x_2 \;\delta (x_1-x_2).
\label{P12 1}
\end{eqnarray}
%%%%%%%%%%%%%%%%%%%%%%%%%%%
We write, for the above exponential, the Taylor expansion
%%%%%%%%%%%%%%%%%%%%%%%%%%%%%%%
\begin{eqnarray}
e^{i\left[ \phi_{\theta}(x_2)-\phi_{\theta}(x_1)\right] }
&=& 1 + i\left[(x_2-x_1)\phi'_{\theta}(x_1) + \frac{(x_2-x_1)^2}{2!}\phi''_{\theta}(x_1)
+ \cdots\right]
\nonumber \\
&&-\frac{1}{2!}\left[(x_2-x_1)^2(\phi'_{\theta}(x_1))^2 + 
2\frac{(x_2-x_1)^3}{2!}\phi'_{\theta}(x_1)\phi''_{\theta}(x_1)+ \cdots  \right]
\nonumber \\
&&+\cdots \; ,
\label{taylor exponential}
\end{eqnarray}
%%%%%%%%%%%%%%%%%%%%%%%%%%%}
where the primes mean derivatives with respect to the argument.
We use the $\delta$-function identities
%%%%%%%%%%%%%%%%%%%%%%%%%%%%%%%
$
x \delta '(x) = - \delta (x)
$,
$
x^n \delta '(x) = 0, \; n\ge 2 
$,
%%%%%%%%%%%%%%%%%%%%%%%%%%%
to write the matrix element, Eq. (\ref{P12 1}), as 
%%%%%%%%%%%%%%%%%%%%%%%%%%%%%%%
\begin{eqnarray}
\langle x_1, \theta |\hat{P}_{\theta}| x_2, \theta \rangle
= \left[-i \delta ' (x_1-x_2) + \phi'_{\theta} (x_1)
\delta (x_1-x_2)\right]
+\frac{\cos \theta}{\sin \theta} x_1 \delta (x_1-x_2).
\label{P12 1a}
\end{eqnarray}
%%%%%%%%%%%%%%%%%%%%%%%%%%%
In order to satisfy the second requirement in Eq. (\ref{m.els. X and P}) we thus need
%%%%%%%%%%%%%%%%%%%%%%%%%%%%%%%
\begin{equation}
\phi'_{\theta}(x') = - \frac{\cos \theta}{\sin \theta} x' \;,
\end{equation}
%%%%%%%%%%%%%%%%%%%
with the solution
%%%%%%%%%%%%%%%%%%%%%%%%%%%%%%%
\begin{equation}
\phi_{\theta}(x') = - \frac{\cos \theta}{\sin \theta}\frac{x'^2}{2} + 
\varphi(\theta),
\end{equation}
%%%%%%%%%%%%%%%%%%%
where $\varphi(\theta)$ is an arbitrary function of $\theta$.

The wave function $\psi_{x', \theta}(x)$ of 
Eq. (\ref{psi_(x,theta)(x') 3}) then becomes
%%%%%%%%%%%%%%%%%%%%%%%%%%%%%%%%%
\begin{equation}
\psi_{x', \theta}(x)
= \frac{e^{i\varphi(\theta)}}{\sqrt{2 \pi |\sin \theta|}} \;
e^{-\frac{i}{2 \sin \theta}[(x^2+x'^2)\cos \theta - 2x x']} \; .
\label{psi_(x,theta)(x') 4}
\end{equation}
%%%%%%%%%%%%%%%%%%%%%%%%%%%%%%%%

4) Choosing, for the phase $\varphi(\theta)$, 
%%%%%%%%%%%%%%%%%%%%%%%%%%%%%%%%%
\begin{equation}
\varphi(\theta) = \frac{\pi}{4}{\rm sgn}(\sin\theta) -\frac{\theta}{2},
\label{phi(theta)}
\end{equation}
%%%%%%%%%%%%%%%%%%%%%%%%%%%%%%%%
we satisfy the requirements of Eq. (\ref{lim theta to 0 and plane_wave}).

We finally find the wave function of Eq. (\ref{psi_{x,theta}(x') 2}).

%%%%%%%%%%%%%%%%%%%%%%%%%%%%%%%%%%%%%
\section{The matrix elements of the operator $\exp(i\theta \hat{n})$}
\label{m.els.exp(itn)}

In this Appendix we compute the matrix elements of the operator $\exp{(i\theta \hat{n})}$ in the original basis $|x \rangle$.
We have
%%%%%%%%%%%%%%%%%%%%%%%%%%%%%%%%%%%%%%%%
\begin{subequations}
\begin{eqnarray}
\langle x|e^{i\theta \hat{n}}|x' \rangle 
&=& \sum_n \psi_n^*(x)\psi_n(x')e^{in\theta}   
\label{x' U+ x a} \\
&=& \frac{1}{\sqrt{\pi}}e^{-\frac{x^2+x'^2}{2}}
\sum_n H_n(x)H_n(x')\frac{(e^{i\theta})^n}{2^n n!},
\label{x' U+ x b}
\end{eqnarray}
\end{subequations}
%%%%%%%%%%%%%%%%%%%%%%%%%%%%%%%%%%%%%%%%
where $\psi_n(x)$ are the one-dimensional harmonic oscillator wave functions
(see Ref. \cite{merzbacher}, p. 61, Eq. (5.24), where the variable $x$ has been replaced by the dimensionless $x/\sqrt{\frac{\hbar}{m \omega}}$, as used in this paper)
%%%%%%%%%%%%%%%%%%%%%%%%%%%%%%%%%
\begin{equation}
\psi_n(x)
= \frac{1}{\pi^{1/4}\sqrt{2^n n!}}e^{-\frac{x^2}{2}} H_n(x),
\label{ho-wf}
\end{equation}
%%%%%%%%%%%%%%%%%%%
$H_n(x)$ being Hermite polynomials.
We compute the sum in Eq. (\ref{x' U+ x b}) using the identity
(Ref. \cite{morse-feshbach}, p. 781, Problem 6.12)
%%%%%%%%%%%%%%%%%%%%%%%%%%%%%%%%%
\begin{equation}
\sum_n H_n(x) H_n(x')\frac{t^n}{2^n n!}
= \frac{1}{\sqrt{1-t^2}}\exp\left[{\frac{2x x' t - t^2 (x^2 + x'^2)}{1-t^2}}\right],
\label{morse-feshbach identity}
\end{equation}
%%%%%%%%%%%%%%%%%%%
with the result
%%%%%%%%%%%%%%%%%%%%%%%%%%%%%%%%%
\begin{equation}
\langle x|e^{i\theta \hat{n}}|x' \rangle 
= e^{-i\frac{\theta}{2}} \;
\frac{e^{-\frac{i}{2 \sin \theta}[(x^2+x'^2)\cos \theta - 2x x']}}
{\sqrt{2 \pi (-i) \sin \theta}} .
\label{x' U+ x 2}
\end{equation}
%%%%%%%%%%%%%%%%%%%
This result is identical to that of Eq. (\ref{<x'U+x>}) if we choose, for the square root, the branch
%%%%%%%%%%%%%%%%%%%%%%%%%%%%%%%%%
\begin{eqnarray}
\sqrt{-i \sin \theta} 
&=&
\left\{
\begin{array}{c}
e^{-i\frac{\pi}{4}} \sqrt{|\sin \theta|}, \;\;\; {\rm for} \;\;\; \sin \theta > 0        \\
e^{+i\frac{\pi}{4}} \sqrt{|\sin \theta|}, \;\;\; {\rm for} \;\;\; \sin \theta < 0 
\end{array}
\right. ,
\\
&=&e^{-i\frac{\pi}{4}{\rm sgn}(\sin \theta)}
\sqrt{|\sin \theta|}.
\end{eqnarray}
%%%%%%%%%%%%%%%%%%%

%%%%%%%%%%%%%%%%%%%%%%%%%
\section{A simple way to derive the result (\ref{|<x'theta'|x theta>|})}
\label{graphical_method}

\begin{figure}[h]
\epsfxsize=0.3\textwidth
\epsfysize=0.3\textwidth  
\centerline{\epsffile{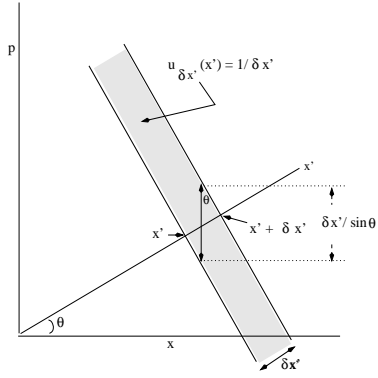}}
\caption{
\footnotesize{
Geometric argument to arrive at Eq. (\ref{|<x|x' theta>|2 3}).
In the shaded area, whose width is $\delta x'$, 
$u_{\delta x'} (x') =  1/\delta x'$.}}
\label{geom_inter_S}
\end{figure}

It will suffice to evaluate the quantity
$|\langle x|x';\theta \rangle|^2$;
this is the probability to find $x$ in a unit interval around the value $x$ when the system has been prepared in the state 
$|x';\theta \rangle$. We find
%%%%%%%%%%%%%%%%%%%%%%%%%%%%%%%%%%
\begin{subequations}
\begin{eqnarray}
|\langle x|x';\theta \rangle|^2
&=& \langle x |\hat{\mathbb{P}}_{x';\theta}|x \rangle
=Tr(\hat{\mathbb{P}}_{x';\theta} \hat{\mathbb{P}}_x)
\label{|<x|x' theta>|2 a} \\
&=& \int \int W_{\hat{\mathbb{P}}_{x';\theta}} (q,p)
W_{\hat{\mathbb{P}}_{x}}(q,p)
\frac{dq dp}{2 \pi} \; ,
\label{WT |<x|x' theta>|2 b}
\end{eqnarray}
\label{|<x|x' theta>|2 1}
\end{subequations}
%%%%%%%%%%%%%%%%%%%%%%%%%%%%%%%%
where in the last line we have used Eq. (\ref{Tr(AB)}) to express our probability in terms of Wigner transforms.
The Wigner transform of the projector $\hat{\mathbb{P}}_{x';\theta}$    is found from Eq. (\ref{WT_P_x'_theta}) and that for the projector 
$\hat{\mathbb{P}}_{x}$ is simply $\delta(x-q)$. 
We thus write the last equation as
%%%%%%%%%%%%%%%%%%%%%%%%%%%%%%%%%%
\begin{subequations}
\begin{eqnarray}
|\langle x|x';\theta \rangle|^2
&=& \int \int 
\delta (x' - (q C + p S)) \delta(x-q)
\frac{dq dp}{2 \pi} \\
&=& \frac{1}{2 \pi} \int \delta (x'-(x C + p S)) dp 
\label{|<x|x' theta>|2 2 b} \; .
\end{eqnarray}
\label{|<x|x' theta>|2 2}
\end{subequations}
%%%%%%%%%%%%%%%%%%%%%%%%%%%%%%%%
On the one hand, this integral can be evaluated directly, giving
%%%%%%%%%%%%%%%%%%%%%%%%%%%%%%
\begin{equation}
|\langle x|x';\theta \rangle|^2
=\frac{1}{2 \pi} \int \frac{1}{|S|}
\delta\left (p-\frac{x'-xC}{S} \right) dp 
=\frac{1}{2 \pi |S|} \; ,
\label{|<x|x' theta>|2 3}
\end{equation}
%%%%%%%%%%%%%%%%%%%%%%%%%%%%%%
just as in Eq. (\ref{|<x'theta'|x theta>|}) with $\theta_1=\theta$ and
$\theta_2=0$.
On the other hand, the appearance of the factor $|S|$ in the denominator of the result (\ref{|<x|x' theta>|2 3}) can be understood by using an intuitive geometrical argument starting from (\ref{|<x|x' theta>|2 2 b}), as follows.
We approximate the delta function occurring in Eq. (\ref{|<x|x' theta>|2 2 b})
by the step
%%%%%%%%%%%%%%%%%%%%%%%%%%%%%%
\begin{equation}
\delta (x'-(x C + p S))
\approx u_{\delta x'} (x')
\equiv 
\left\{
\begin{array}{c}
\frac{1}{\delta x'}, \;\;\; {\rm if}\;\;\;   
x'\in (x'-\frac{\delta x'}{2}, x'+\frac{\delta x'}{2}) \\
0, \;\;\; {\rm if}\;\;\;
x' \notin (x'-\frac{\delta x'}{2}, x'+\frac{\delta x'}{2}) 
\end{array}
\right. \; ,
\label{step}
\end{equation}
%%%%%%%%%%%%%%%%%%%%%%
the delta function being attained in the limit $\delta x' \to 0$.
The non-zero region is indicated as the shaded area in Fig. \ref{geom_inter_S}.
The segment along the $p$-axis, over which we are integrating, contained inside that area, is $\delta p = \delta x' / \sin \theta$. 
The integral in Eq. (\ref{|<x|x' theta>|2 2 b}) is thus given by 
%%%%%%%%%%%%%%%%%%%%%%%%%%%%%%
\begin{equation}
\frac{1}{\delta x'} \; \frac{\delta x'}{\sin \theta} 
=  \frac{1}{\sin \theta}        \; ,
\label{|<x|x' theta>|2 4}
\end{equation}
%%%%%%%%%%%%%%%%%%%%%%
thus reproducing the formal result of Eq. (\ref{|<x|x' theta>|2 3}).
%%%%%%%%%%%%%%%%%%%%%%%%%%%%%%%%%%%%%%
%%%%%%%%%%%%%%%%%%%%%%%%%%%%%%%%%%%%%%

%%%%%%%%%%%%%%%%%%%%%%%%%%%%

\end{document}